\documentclass[usenatbib]{mn2e}

\usepackage{times}
\usepackage{lscape}
\usepackage{longtable}
\usepackage{graphicx}
\usepackage{float}
\usepackage{morefloats}
\usepackage{subfigure}

\author[S. Alaghband-Zadeh et al.]
{S.\ Alaghband-Zadeh,$^1$ S.\ C.\ Chapman,$^1$ A.\ M.\ Swinbank,$^2$ Ian Smail,$^2$ C.\ M.\ Harrison,$^3$
\newauthor
D.\ M.\ Alexander,$^3$ C.\ M.\ Casey,$^4$ R.\ Dav\'{e},$^5$  D.\ Narayanan,$^5$ Y.\ Tamura,$^6$ H.\ Umehata$^6$\\
$^{1}${Institute of Astronomy, Madingley Road, Cambridge, CB3 0HA UK}\\
$^{2}${Institute for Computational Cosmology, Durham University, South Rd, Durham DH1 3LE, UK}\\
$^{3}${Department of Physics, Durham University, South Rd, Durham DH1 3LE, UK}\\
$^{4}${Institute for Astronomy, 2680 Woodlawn Dr, Honolulu, HI 96822, USA}\\
$^{5}${Astronomy Department, University of Arizona, Tucson, AZ 85721, USA}\\
$^{6}${Institute of Astronomy, The University of Tokyo, 2-21-1 Osawa, Mitaka, Tokyo 181-0015, Japan}\\
}

\date{\today}
\begin{document}

\title{Integral Field Spectroscopy of 2.0$<$z$<$2.7 Sub-mm Galaxies; gas morphologies and kinematics}
\maketitle
\begin{abstract}
We present two-dimensional, integral field spectroscopy covering the rest-frame wavelengths of strong optical emission lines in nine sub-mm-luminous galaxies (SMGs) at 2.0$<$z$<$2.7. The GEMINI-NIFS and VLT-SINFONI imaging spectroscopy allows the mapping of the gas morphologies and dynamics within the sources, and we measure an average H$\alpha$ velocity dispersion of $\langle\sigma\rangle=220\pm80$\,kms$^{-1}$ and an average half light radius of $\rm \langle r_{1/2}\rangle=3.7\pm0.8$\,kpc. The dynamical measure, $\langle \rm V_{obs}/2\sigma \rangle=0.9\pm0.1$ for the SMGs, is  higher than in more quiescent star-forming galaxies at the same redshift, highlighting a difference in the dynamics of the two populations. The far-infrared SFRs of the SMGs, measured using {\it Herschel}-SPIRE\footnotemark far-infrared photometry, are on average 370$\pm$90\,$\rm M_{\odot}yr^{-1}$ which is $\sim$2 times higher than the extinction corrected SFRs of the more quiescent star-forming galaxies. Six of the SMGs in our sample show strong evidence for kinematically distinct multiple components with average velocity offsets of 200$\pm$100\,kms$^{-1}$ and average projected spatial offsets of 8$\pm$2\,kpc, which we attribute to systems in the early stages of major mergers. Indeed all SMGs are classified as mergers from a kinemetry analysis of the velocity and dispersion field asymmetry. We bring together our sample with the seven other SMGs with IFU observations to describe the ionized gas morphologies and kinematics in a sample of 16 SMGs. By comparing the velocity and spatial offsets of the SMG H$\alpha$ components with sub-halo offsets in the Millennium simulation database we infer an average halo mass for SMGs in the range of $\rm 13<log(M[h^{-1}M_{\odot}])<14$. Finally we explore the relationship between the velocity dispersion and star formation intensity within the SMGs, finding the gas motions are consistent with the Kennicutt-Schmidt law and a range of extinction corrections, although might also be driven by the tidal torques from merging or even the star formation itself.
\end{abstract}

\footnotetext{Herschel is an ESA space observatory with science instruments provided by European-led Principal Investigator consortia and with important participation from NASA.}

\section{Introduction}
\label{sec:intro}

Studies of the H$\alpha$ nebular emission line have revolutionized our understanding of moderate luminosity star forming galaxies at z$\sim$2.  The most extensive study to date has been through the Spectroscopic Imaging survey in the Near-infrared with SINFONI (SINS) of 62 star-forming galaxies at $1.3<z<2.6$ \citep{Forster-Schreiber09}, a subset of known H$\alpha$ emitters from the UV-continuum selected galaxies \citep{Steidel04,Daddi04}. In the SINS galaxies, the H$\alpha$ velocity fields are mapped to trace the dynamics of the ionized gas within the galaxies, with a striking result that the star-forming galaxies can be split into three equally-populated groups; turbulent disks which are rotation dominated, compact sources which are dispersion dominated, and merging systems. These galaxies populate the `main sequence' of star-forming galaxies at these redshifts \citep[eg][]{Daddi07,Noeske07}. The SINS survey, however,  only has 6\% of galaxies with inferred star-formation rates (SFRs) greater than 500\,M$_{\odot}$\,yr$^{-1}$, which is the domain of galaxies selected through bright far-infrared emission redshifted into the sub-millimetre wavelengths, or sub-millimetre galaxies (SMGs). Further, the poorly constrained dust extinction corrections in the SINS galaxies implies that this `high-SFR' tail may actually have substantially lower SFRs. There is thus an obvious need to explore the nebular emission line dynamics of the SMG population to compare to the `main sequence' of star-forming galaxies. 

Since their discovery, SMGs have been the focus of intense multi-wavelength study. With 15 years since the first sub-mm surveys \citep[eg][]{Smail97, Hughes98} our understanding of this population has grown considerably, although still limited by only handfuls of objects studied in detail at high spatial resolution \citep[eg][]{Swinbank06,Engel10,Carilli10}. These dusty galaxies have a redshift distribution which peaks at $z\sim$2.5 \citep{Chapman05,Wardlow11}. At these redshifts, the observed sub-mm flux densities imply far-infrared luminosities $>$10$^{12}$\,L$_{\odot}$ (Ultra-Luminous Infra-Red Galaxies -- ULIRGs).  With space densities at $z\sim$2 a factor $\sim$1000$\times$ higher than today \citep{Chapman03}, these luminous systems contribute $\sim$30\% of the star-formation activity at $z\sim$2--3, compared to $<$1\% locally. It is noteworthy that some confirmed ULIRGs at $z\sim$2 have hotter SED shapes resulting in 850\,$\mu$m fluxes lying below the sensitivities of current surveys, and the high-redshift  ULIRG population could be up to twice as large as implied by 850\,$\mu$m surveys alone  \citep{Chapman04a,Casey09a,Magdis10}.  These galaxies which have been missed by 850\,$\mu$m surveys, have been detected by BLAST (\citealt{Casey11}) and {\it Herschel} (\citealt{Magnelli10,Chapman10}), and confirmed as `SMGs' with 250--500\,$\mu$m flux densities still implying ULIRG-like luminosities. The intense star-formation rates (SFRs$\sim$1000M$_{\odot}$\,yr$^{-1}$) and large molecular gas reservoirs ($\sim$10$^{11}$\,M$_{\odot}$; \citealt{Greve05}; Bothwell et al.\ in prep.) inferred for the SMGs means that they could build a stellar mass of 10$^{11}$\,M$_{\odot}$ in just 100\,Myr. The space densities and inferred duty cycles of SMGs suggest they represent the formation phase of today's massive elliptical galaxies \citep{Lilly99, Genzel03, Swinbank08}.  

SMGs are clearly an influential population of early galaxy evolution, but attempts to model their basic properties have proved problematic. The most successful semi-analytic models have had to radically alter their prescriptions for starbursts during mergers in order to account for SMGs whilst remaining self-consistent with other constraints \citep{Baugh05}. \cite{Narayanan09} looked at isolated conditions in hydrodynamical simulations that could lead to very high SFRs, and proposed a merger-driven formation scenario for SMGs. Subsequent analysis by \cite{Hayward11b} showed that these merging galaxies may represent the bright end of the SMG number counts, but found that the fainter tail of SMGs ($S_{850}\sim$3\,mJy) are typically made up of isolated galaxies and galaxy pairs. \cite{Dave10} identified SMGs in their cosmological simulations as typically very massive galaxies which are being harassed by small fragments, offering an alternative possibility to the major-merger picture for SMGs. Observations of individual SMGs at high-resolution are therefore required to guide the models and constrain the driver for the high SFRs.

High-resolution radio and CO observations have shown that the star-formation and gas can be extended on scales up to $\sim10$\,kpc  in SMGs, and multiple components are common within these extended distributions \citep{Chapman04b,Biggs08,Tacconi08,Casey09a,Bothwell10,Ivison11}. These sizes are much larger than the typical $\sim$1\,kpc extent of local ULIRGs \citep[eg][]{Sakamoto08} whose intense SFRs are often triggered by mergers and interactions \citep{Sanders96}. \cite{Engel10} and \cite{Bothwell10} concluded that most SMGs are formed in major mergers, many exhibiting distinct kinematic components and disturbed gas morphologies (indicative of pre-coalesced to late stage mergers) from their high resolution interferometric CO line studies.

Studying SMGs in H$\alpha$ complements the CO studies, as lower levels of star formation are detectable through H$\alpha$ emission than through the existing CO or radio measurements  (from the IRAM-Plateau de Bure Interferometer and the Very Large Array respectively). Further, merging galaxies may not induce similarly high levels of star formation in all components of the merger, and H$\alpha$ may elucidate a more extensive merging system than apparent in CO. \cite{Swinbank06} find multiple components in four of the six SMGs they studied using integral field unit (IFU) spectroscopy to map H$\alpha$, with average velocity offsets of $\sim$180\,kms$^{-1}$ and spatial offsets up to $\sim$8\,kpc. They concluded that SMGs are similar to the local, more compact ULIRGs in that they exhibit multiple components. However, they were not able to resolve kinematic information on scales $<$5\,kpc  (i.e., within each component) and thus the gas dynamics of the multi-component systems could not be established. Using higher spatial resolution IFU observations, \cite{Menendez-Delmestre12}  measure the gas kinematics within the multiple components of three SMGs.  However the majority of the H$\alpha$ fields of these three sources are dominated by broad-line AGN, and only small sections of the narrow line regions have [NII]/H$\alpha$ ratios which are consistent with photo-ionization from star formation. There has thus remained a need to further explore a larger sample of SMGs with high resolution, sensitive IFU spectroscopy, including the large fraction of SMGs ($>$50\%) which do not show such obvious indications of strong AGN in their optical spectra.

In this paper we present the gas dynamics of nine SMGs observed with the Gemini-NIFS and VLT-SINFONI. In Section \ref{sec:obs} we present the sample and the observations.  In Section \ref{sec:analysis} we analyse the spatially integrated and resolved observations. The dynamical properties of the SMGs are presented in Section \ref{sec:results} along with the comparison to other galaxy surveys and the kinemetry analysis.  In Section \ref{sec:discuss} we discuss the merger properties and use the component properties to infer the average halo mass. Finally, in Section \ref{sec:conc} we give our main conclusions. All calculations assume a flat, $\Lambda$CDM cosmology with $\Omega_\Lambda=0.7$ and $H_0=71$\,kms$^{-1}$Mpc$^{-1}$ in which 0.5$''$ corresponds to a physical scale of $\sim$4\,kpc at $z=2$ (the median seeing and redshift of our observations and targets respectively).

\section{Sample selection and Observations}
\label{sec:obs}
\subsection{Sample}

Our IFU survey comprises nine radio-identified ULIRGs (listed in Table \ref{tab:ref}), targeted from various surveys, preferentially chosen from our equatorial and southern fields to enable further study with ALMA.  {\it Herschel}-SPIRE fluxes are available for the majority of the sources (Table \ref{tab:ref}) enabling the far-infrared luminosities of the samples to be derived directly. We find that all sources have $\rm L_{FIR}>$10$^{12}$\,L$_{\odot}$. The targets designated `SMM'  are SMGs with $S_{850\mu m}>4$\,mJy, while those listed as `RG' are undetected at 850\,$\mu$m, with typical 3$\sigma$ rms of 4.5\,mJy. These sources are, however, still found to be bright at 250, 350, \& 500\,$\mu$m and therefore represent the ULIRGs with hotter dust temperatures causing the  850\,$\mu$m fluxes to fall below the detection limits. Eight of the sources have spectroscopic redshifts derived from our longslit detections of the H$\alpha$ emission line. For the ninth source, SMMJ2217+0017, we obtained the redshift directly with our IFU observations. SMMJ2217+0017 was identified as a bright SMG ($S_{1100\mu m}=4.9\pm0.7$\,mJy) in the SA22 field \citep{Tamura09} with a VLA counterpart \citep{Chapman04c}, and an optical through mid-IR ({\it Spitzer-}IRAC) photometric redshift constrained to $z\sim$2--2.5.

\begin{figure}
\includegraphics[width=0.5\textwidth]{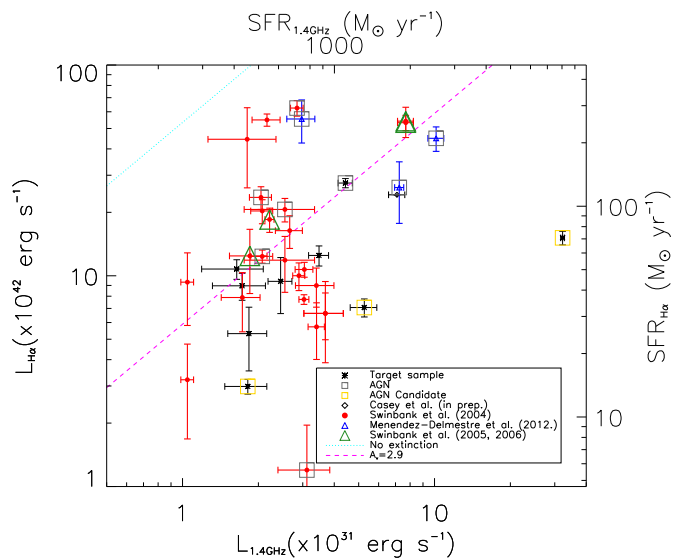}
\caption{The H$\alpha$ luminosities from our IFU observations (not corrected for extinction) versus radio luminosities of the sources studied in this work, compared to the sample of SMGs studied in \protect \cite{Swinbank04}using long-slit spectroscopy. The $\rm SFR_{1.4GHz}$ axis represents the SFRs derived from the radio luminosities, using the far-infrared-radio correlation \protect \citep{Ivison10} and the conversion to SFR from \protect \cite{Kennicutt98b}. The $\rm SFR_{H\alpha}$ axis assumes the conversion to SFR from \protect \cite{Kennicutt98b} and both $\rm SFR_{1.4GHz}$ and $\rm SFR_{H\alpha}$ conversions include a factor of 1.7 to convert from the Salpeter to Chabrier IMF. The sample of SMGs with previously published IFU observations \protect \citep{Swinbank05,Swinbank06,Menendez-Delmestre12} are highlighted. We also plot the H$\alpha$ luminosity for RGJ0332$-$2758 \protect \citep{Casey11} which has IFU observations presented in Casey et al.\ (in prep.). Our target sample has a similar range of radio and H$\alpha$ luminosities to the SMGs in \protect \cite{Swinbank04}. The dotted line denotes equality in the two SFR indicators whereas the dashed line correspond to the relationship between the two SFR indicators for an extinction of $\rm A_v=2.9\pm0.5$ (the average value for SMGs from \protect \citealt{Takata06}) which decreases the observed SFR$_{H\alpha}$ relative to the radio. We also mark the target sources which potentially host AGN, as discussed in Section 3.3 and the previously published sources which are known to host AGN.}
\label{fig:selec}
\end{figure}

The SMGs span a redshift range of 2.0$<z<$2.7, which covers the median of the parent sample from \cite{Chapman05}, and places H$\alpha$ in the $K$-band atmospheric window. Fig. \ref{fig:selec} shows the  range of 1.4\,GHz radio (referenced in Table \ref{tab:ref}) and H$\alpha$ luminosities in the target sample compared to the larger sample of SMGs studied in \cite{Swinbank04} and \cite{Chapman05},  demonstrating our IFU sample is representative. We use the radio luminosities and radio-derived SFRs in Fig. \ref{fig:selec}, rather than the directly measured far-infrared luminosities and $\rm SFR_{FIR}$ values, to enable a comparison to the complete parent sample. Fig. \ref{fig:selec} also shows that our sample is consistent with having extinctions of A$_v\sim$2.9, corresponding to the average A$_v$ measured for a sample of SMGs from the H$\alpha$/H$\beta$ ratio \citep{Takata06}, which brings the H$\alpha$ SFRs in line with those inferred from the far-infrared.

\begin{table*}
\centering
\begin{tabular}{|l|c|c|c|c|c|c|c|c|c|c|}
\hline\hline
Source &  RA & Dec & z$_{\rm H\alpha}$ & $S_{\rm 1.4GHz}$ & $S_{\rm 250\mu m}$  & $S_{\rm 350\mu m}$  & $S_{\rm 500\mu m}$ & $S_{\rm 850\mu m}$ & $\rm t_{int}$ & Notes \\
\hline
 & (J2000) & (J2000) & & ($\mu$Jy) & (mJy) & (mJy) & (mJy) & (mJy) & (ks) & \\
\hline\hline
RGJ0331$-$2739$^1$    &  03:31:52.82 & $-$27:39:31.5   &  2.3429[1] & $965\pm16$   & $36.0\pm3.6$ & $47.0\pm4.2$ & $42.0\pm4.8$  & $1.9\pm1.5$ & 4.8 & 1 \\
SMMJ0217$-$0505       &  02:17:38.92 & $-$05:05:23.7  &  2.5305[2] & $41\pm11$     & $24.8\pm3.6$ & $26.2\pm4.2$ & $26.0\pm4.8$  & $7.1\pm1.5$ & 4.8 & 2\\    
SMMJ0333$-$2745       &  03:33:15.43 & $-$27:45:24.4 &  2.6937[2] & $75.8\pm6.9$   & $31.2\pm3.6$ & $41.0\pm4.2$ & $35.3\pm4.8$  & $9.2\pm1.2$  & 4.8  & 3\\ 
SMMJ2218+0021       &  22:18:04.42 & 00:21:54.4    &  2.5171[4] & $44\pm10$        &  $-$ & $-$  &  $-$  & $9.0\pm2.3$  & 4.2  & 4\\   
SMMJ0332$-$2755       &  03:32:43.20 & $-$27:55:14.5   &  2.1236[3] & $106\pm9$    & $32.8\pm3.6$ & $31.6\pm4.2$ & $21.2\pm4.8$  & $5.2\pm1.4$  & 3.6  & 5\\   
SMMJ2217+0017$^1$   &  22:17:42.25 & 00:17:02.0    &  2.2777[2] & $58\pm11$        & $-$ & $-$  &   $-$ &  $-$ & 9.6  & 6\\  
RGJ0332$-$2732$^1$    &  03:32:56.75  & $-$27:32:06.3  &  2.3152[3] & $160\pm 20$  & $27.5\pm3.6$ & $27.5\pm4.2$ & $20.3\pm4.8$  & $2.4\pm1.4$  & 2.4  & 7\\
SMMJ0217$-$0503$^2$   &  02:17:38.62 &	$-$05:03:39.5   &  2.0280[3] & $185\pm 12$ & $36.3\pm3.6$ & $33.9\pm4.2$ & $21.7\pm4.8$  & $4.4\pm1.7$  & 4.8  & 2\\  
SMMJ0217$-$0459       &  02:17:25.12 &	$-$04:59:37.4   &  2.3090[5] & $57\pm10$   & $21.7\pm3.6$ & $22.2\pm4.2$ & $17.0\pm4.8$  & $4.5\pm1.9$  & 7.2  & 2\\ 
\end{tabular}
\caption{ Table of the SMG fluxes, positions, redshifts and on source exposure times. The redshifts quoted are derived from the IFU spectra measurements of the H$\alpha$ emission line. The values in [] in the redshift column represent the error on the last decimal place. The radio flux and position references are given in the final column.
1: \protect \cite{Ivison10}, 
2: \protect \cite{Ivison07},
3: \protect \cite{Biggs11},
4: \protect \cite{Chapman05},
5: \protect \cite{Dunlop10},
6: \protect \cite{Chapman04c},
7: \protect \cite{Miller08}. $^1$AGN candidate, and $^2$broad line AGN, as discussed in Section 3.3. The $S_{\rm 250\mu m}$, $S_{\rm 350\mu m}$  and $S_{\rm 500\mu m}$ fluxes are measured from {\it Herschel}-SPIRE observations \protect \citep{Griffin10,Oliver12}. The errors quoted are the source extraction errors but we note these are dominated by the confusion noise (5.8, 6.3 and 6.8\,mJy/beam for $S_{\rm 250\mu m}$, $S_{\rm 350\mu m}$ and $S_{\rm 500\mu m}$ respectively). The $S_{\rm 850\mu m}$ fluxes are taken from \protect \cite{Chapman05,Coppin06,Weiss09}. }
\label{tab:ref}
\end{table*}

\subsection{Observations and Reduction}

Observations of RGJ0331$-$2739, SMMJ0217$-$0505, SMMJ0333$-$2745, SMMJ2218+0021 and SMMJ0332$-$2755  were made with the Gemini-North/Near-Infrared Integral Field Spectrograph (NIFS) between 2010 September 24th and 2011 February 28th as part of program GN/2010B/42. The Gemini-NIFS IFU uses an image slicer to take a 3.0$''\times$3.0$''$ field and divides it into 29 slices of width 0.103$''$. The dispersed spectra from the slices are reformatted on the detector to provide 2-dimensional spectro-imaging, in our case using the $K$-band grism covering a wavelength range of 2.00--2.43\,$\mu$m. The observations were performed using an ABBA configuration in which we chopped by 10$''$ to blank sky to achieve sky subtraction. Individual exposures were 600\,s and each observing block was 2.4\,ks, which was repeated between three and four times resulting in the on source integration times (which do not include the sky observations) given in Table \ref{tab:ref}.

SMMJ2217+0017, SMMJ0217$-$0503, RGJ0332$-$2732 and SMMJ0217$-$0459 were observed with the Spectrograph for INtegral Field Observations in the Near Infrared (SINFONI) on the VLT between 2011 July 13th and 2011 September 23rd as part of programme 087.A-0660(A). All of our observations targeted the redshifted H$\alpha$ and [N{\sc ii}]$\lambda\lambda$6548.1,6583.0 emission lines in the $K$-band. We use the $H+K$ grating on SINFONI which gives a spectral resolution R=1500 covering a range of 1.45$–-$2.45\,$\mu$m. The field of view is sliced into 32 slices each of which is imaged onto 64 pixels. We chose the width of the slices to be 0.25'' to  provide a 8$''\times$8$''$ field (obtaining 32 by 64 spectra of the image with each pixel covering 0.25'' by 0.125''). We move the target around 4 quadrants of the IFU for sky subtraction purposes, however retaining the target in the field of view, therefore the effective field of view is 4$''\times$4$''$ for SMMJ2217+0017, RGJ0332$-$2732 and SMMJ0217$-$0459  and 4$''\times$8$''$ for SMMJ0217$-$0503 where we only use 2 quadrants in order to observe the entire extended system. The individual exposure times are 600s with resulting on source exposure times (which do not include the sky observations) given in Table \ref{tab:ref}.

We reduced the NIFS data with the standard Gemini {\sc iraf nifs} pipeline  and the SINFONI data with the SINFONI ESOREX pipelines both of which include extraction, sky subtraction, wavelength calibration, and flat-fielding.  To accurately align and mosaic the individual datacubes we created white light (wavelength collapsed) images around the redshifted H$\alpha$ emission line from each observing block and centroid the galaxy within the data cube.  These were then spatially aligned and co-added using an average with a 3$\sigma$ clipping threshold to remove remaining cosmetic defects and cosmic rays. Flux calibration was carried out by observing bright A0V standard stars at similar airmass to the target galaxy immediately after each observing block for both the NIFS and SINFONI observations, and we then also compare the integrated fluxes to those measured from longslit observations of the targets. From the reduced standard star cubes, we measure a seeing of FWHM$\sim$0.5$''$ for both the NIFS and SINFONI observations which corresponds to $\sim$4\,kpc at $z\sim2$. The spectral resolution of the data (measured from the sky-lines at $\sim$2.2\,$\mu$m) is $\sigma$=1.8\,\AA\ or 25\,kms$^{-1}$ for the NIFS observations and $\sigma$=6.3\,\AA\ or 88\,kms$^{-1}$ for the SINFONI observations. In all following sections, quoted line widths are corrected to take account of the instrumental resolution, $\sigma_{corr}=\sqrt(\sigma_{obs}^2 - \sigma_{sky}^2)$.

\section{Analysis}
\label{sec:analysis}

\subsection{Spatially integrated spectra}
\begin{figure*}
\includegraphics[width=0.95\textwidth]{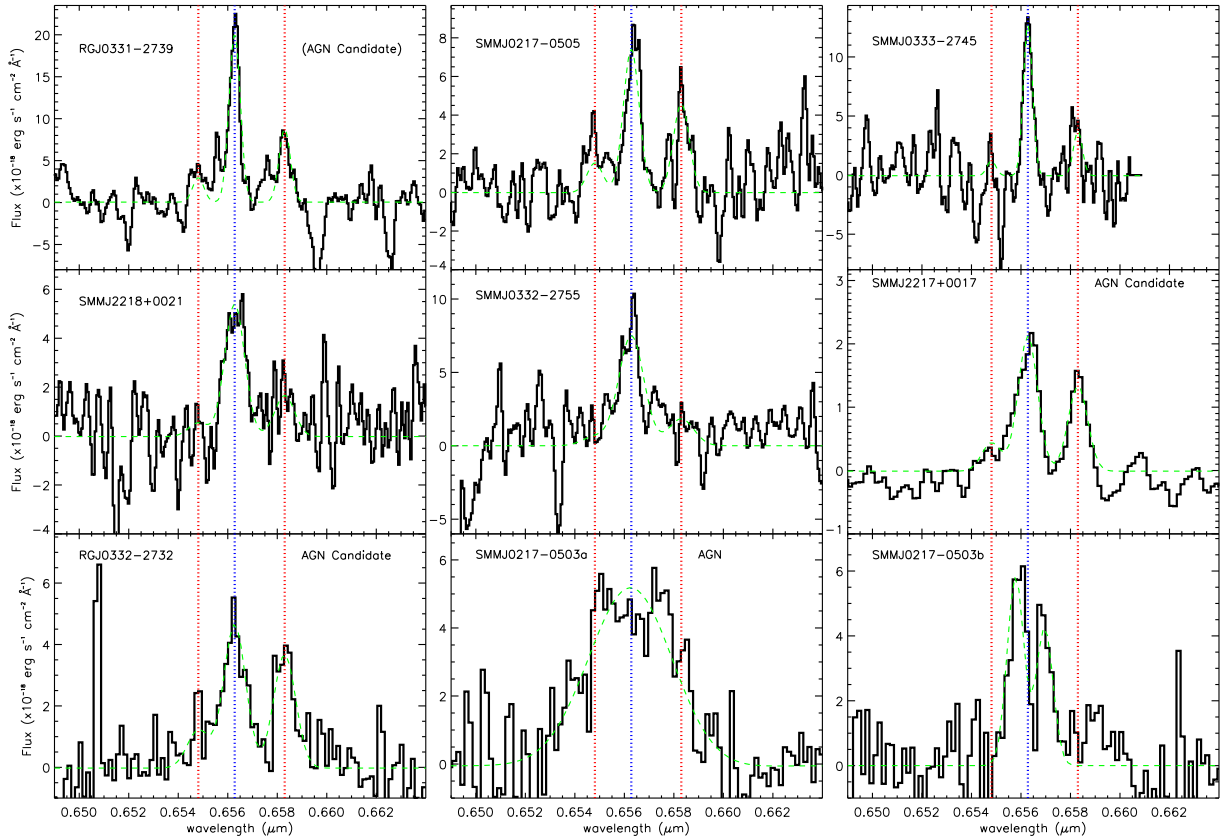}
\caption{1D integrated spectra of sources showing the location of the H$\alpha$ line at 6562.8\,$\mu$m (blue dotted line) and the [NII] lines at 6583.0,6548.1\,$\mu$m (red dotted lines). The NIFS spectra have been smoothed by 2.3 pixels to match the resolution of the SINFONI spectra. All spectra are continuum subtracted. The dashed lines represent the fit to the H$\alpha$ and [NII] lines. The large ratios of [NII]/H$\alpha$ in SMMJ2217+0017 and RGJ0332$-$2732 lead us to suggest that these sources may host AGN. We mark RGJ0331$-$2739 as an `AGN candidate' due to its high radio luminosity, however the low value of [NII]/H$\alpha$ and further analysis of the H$\alpha$ velocity field suggest the H$\alpha$ properties are not being driven by the AGN activity. SMMJ0217$-$0503 is divided into its two clear components (as can be seen in the maps of H$\alpha$). SMMJ0217$-$0503a is most likely an AGN with broad H$\alpha$ emission blending with the [NII] lines. SMMJ0217$-$0503b has a disk-like, double-peaked H$\alpha$ profile.}
\label{fig:spectra}
\end{figure*}

In Fig. \ref{fig:spectra} we show the integrated one-dimensional spectra of the targets with strong H$\alpha$ detections from our IFU observations. We do not include the weak H$\alpha$ detection in SMMJ0217$-$0459 since the IFU observations are low S/N and do not provide further information than the previous longslit observations. We use the integrated spectra to measure basic properties of the sample ($z$, $\rm S_{H\alpha}$ and $\rm \sigma_{H\alpha}$) detailed in Table \ref{tab:basic}. We fit Gaussian profiles to the H$\alpha$  and [NII] emission lines simultaneously in the integrated spectra. We also attempt to fit a broad H$\alpha$ component however we find in all cases that the there is no improvement to the $\chi^2$ of this fit compared to the fit without the broad component so we do not use the broad component fit. Errors on these values were calculated by altering the fit, changing one parameter at a time and allowing the other parameters to minimise, until the $\chi^2$ changed by 1. We only use the flux from the pixels in which H$\alpha$ is detected at $>$3$\sigma$ to create the integrated spectra. For the faint SMMJ0217$-$0459 we quote  the H$\alpha$ redshift (which confirms the redshift gained from previous H$\alpha$ slit observations), H$\alpha$ flux and $\sigma$ but we are unable to measure any resolved properties. For SMMJ0217$-$0503b we fit a double Gaussian to the H$\alpha$ profile since it shows a double-peaked disk-like profile. Indeed the disk-like profile of SMMJ0217$-$0503b is confirmed in Section \ref{sec:kinemetry}.

We measure an average H$\alpha$ flux of $\rm <S_{H\alpha}>=3\pm1\times 10^{-16}$\,erg\,s$^{-1}$\,cm$^{-2}$ and an average velocity dispersion of $\langle\sigma\rangle=220\pm80$\,kms$^{-1}$. In all sources we clearly detect [NII]$\lambda$6583 emission except SMMJ0332$-$2755 for which we quote a 3$\sigma$ upper limit to the ratio of [NII]/H$\alpha$. For our total sample, we measure an average [NII]/H$\alpha$ ratio of 0.52$\pm$0.07 in the sources where [NII] was detected. The large spectral coverage of SINFONI allows for measurements of H$\beta$ in the three SINFONI sources. H$\beta$ however lies in a noise dominated part of the spectra such that no useful limit can be placed on the ratio of H$\alpha$/H$\beta$. We therefore use an extinction value of $\rm A_v=2.9\pm0.5$, taken as the average extinction given in \cite{Takata06}, to correct the H$\alpha$ luminosities since these are dusty systems. The corrected luminosities were converted to SFRs using the conversion from \cite{Kennicutt98a} which includes a conversion factor of 1.7 to adjust from a Salpeter IMF \citep{Salpeter55} to a Chabrier IMF \citep{Chabrier03}. We apply the extinction correction as detailed in \cite{Forster-Schreiber09}, using the correction at the wavelength of H$\alpha$ of  $\rm A_{H\alpha}=0.82A_v$. The total correction applied is therefore $\rm 10^{0.33A_{v}}$. We provide the uncorrected H$\alpha$ fluxes in Table \ref{tab:basic} along with the corrected SFRs.

Since we do not have individual extinction estimates to the H$\alpha$ line, we derive far-infrared luminosities of our SMGs in order to provide unextincted estimates of the SFR. For seven of our SMGs we can use the  {\it Herschel}-SPIRE observations from the DR1 data release \citep{Griffin10,Oliver12} to derive the far-infrared luminosities. We extract the SPIRE fluxes at the positions of our SMGs (Table \ref{tab:ref}) and fit modified black-bodies to the photometry at the known redshift. In this analysis, we fix the dust emissivity, $\beta$\,=\,2 \citep{Magnelli12}, but allow the dust temperature and bolometric luminosity to vary.  For SMMJ2218+0021 and SMMJ2217+0017, for which do not have 250, 350, \& 500\,$\mu$m fluxes, we scale the far-infrared SED of SMM\,J2135-0102 (the well studied lensed SMG at z=2.3; \citealt{Swinbank10a,Ivison11}), to the 850\,$\mu$m flux for SMMJ2218+0021 and the 1.1\,mm flux for SMMJ2217+0017 ($S_{1.1mm}=4.9\pm0.7$\,mJy). We compare the radio luminosities to the far-infrared luminosities to probe the far-infrared radio correlation. We derive, q using Equation \ref{eq:q} \citep{Helou85}.

\begin{equation} 
\rm q= log\left(\frac{L_{FIR}}{9.8 \times 10^{-15} L_\odot}\right) - log\left(\frac{L_{1.4GHz}}{W Hz^{-1}}\right)
\label{eq:q} 
\end{equation}

We find an median q of 2.0$\pm$0.2, for the sources without AGN signatures, and a similar q=1.9$\pm$0.5 for our entire sample, which are consistent with the q value of a sample of SMGs studied in \cite{Kovacs06} (q$\sim$2.1). We convert the derived far-infrared luminosities from the SED fits to SFRs using Equation \ref{eq:sfr}, which is the relation from \cite{Kennicutt98b} adjusted from a Salpeter to a Chabrier IMF using a factor of 1.7. 

\begin{equation} 
\rm SFR(M_\odot\,yr^{-1})= L_{\rm FIR}(ergs^{-1}) \times 2.6\times10^{-44} 
\label{eq:sfr} 
\end{equation}
 
The SFR$\rm_{FIR}$ values are, on average, approximately in line with the extinction corrected SFR$\rm_{H\alpha}$ values indicating that using $\rm A_v=2.9\pm0.5$ brings the two SFR indicators into agreement.

\begin{table*}
%\centering
\scalebox{0.93}{
\begin{tabular}{|l|c|c|c|c|c|c|c|c|c|}
\hline\hline
Source  & S$_{\rm H\alpha}$ & $\rm \sigma_{H\alpha}$ & SFR$_{\rm H\alpha}$  & SFR$_{\rm 1.4GHz}$ & SFR$_{\rm FIR}$  & [NII]/H$\alpha$ & $\rm V_{\rm obs}$  & r$_{1/2}$ & $\rm K_{asym}$  \\
\hline
  & (10$^{-16}$\,erg\,s$\rm^{-1}$\,cm$\rm^{-2}$) & (km\,s$^{-1}$) &  (M$_{\odot}$\,yr$^{-1}$)  & (M$_{\odot}$\,yr$^{-1}$)  &  & (km\,s$^{-1}$) &  & (kpc) &  \\
\hline\hline
RGJ0331$-$2739$^1$ & 3.7$\pm$0.3    &      83$\pm$8  &     600$\pm$200   &   8000$\pm$100  & 280$\pm$60 &  0.41$\pm$0.08  &  120$\pm$20  &  2.99$\pm$0.04  & 0.60$\pm$0.05    \\
SMMJ0217$-$0505 &   2.1$\pm$0.2     &     130$\pm$20 &      500$\pm$200 &   400$\pm$100     & 210$\pm$50 &  0.6$\pm$0.1    &  400$\pm$50      &  2.61$\pm$0.05  & 0.551$\pm$0.007    \\ 
SMMJ0333$-$2745 &   2.1$\pm$0.2     &      70$\pm$10 &     500$\pm$200   &    870$\pm$80  & 340$\pm$50 &  0.3$\pm$0.1    &  150$\pm$30     &  2.05$\pm$0.06  & 5.5$\pm$0.3      \\ 
SMMJ2218+0021  &  1.8$\pm$0.3     &     170$\pm$30 &      400$\pm$200  &    400$\pm$100     & 500$\pm$100  &  0.2$\pm$0.1    &  380$\pm$50     &  2.88$\pm$0.06  & 1.7$\pm$0.1    \\
SMMJ0332$-$2755  &  2.8$\pm$0.8     &     200$\pm$100 &      400$\pm$200  &    610$\pm$60 & 190$\pm$50 &  $<$0.7         &  360$\pm$40  &  2.48$\pm$0.1  & 11$\pm$1       \\
SMMJ2217+0017$^1$  &  0.75$\pm$0.06 &     150$\pm$10 &     130$\pm$50   &    450$\pm$90    & 900$\pm$100 &  0.7$\pm$0.1    &  300$\pm$100  & 4.8$\pm$0.6    & 0.61$\pm$0.05     \\  
RGJ0332$-$2732$^1$  &  1.7$\pm$0.2    &     180$\pm$20 &     300$\pm$100  &    1300$\pm$200 & 200$\pm$40 &  0.8$\pm$0.1    &  200$\pm$100  & 3.1$\pm$0.3    & 1.3$\pm$0.2    \\
SMMJ0217$-$0503 &   9.2$\pm$0.5       &     740$\pm$40 &    1200$\pm$400  &    1110$\pm$70 & 190$\pm$50  &  $-$              &  800$\pm$100  & 9.0$\pm$0.3    & 3.9$\pm$0.3     \\
SMMJ0217$-$0503a$^2$ &  6.4$\pm$0.3   &     740$\pm$40 &     800$\pm$300 &     $-$          & -  &  $-$           &  300$\pm$100  & 4.2$\pm$0.3    & $-$     \\
SMMJ0217$-$0503b &  1.0$\pm$0.1       &     90$\pm$20 &     120$\pm$50   &     $-$         & -  &  0.4$\pm$0.1    &  600$\pm$100  & 4.6$\pm$0.3    & 0.249$\pm$0.005   \\
SMMJ0217$-$0459  &  1.3$\pm$0.4       &     100$\pm$100 &    200$\pm$100    &    460$\pm$80  & 150$\pm$50 &  0.4$\pm$0.3    & $-$              & $-$              & $-$ \\
\end{tabular}
}
\caption{Table of the SMG properties measured and derived from the integrated spectra and maps. The SFR$_{\rm H\alpha}$ values are calculated using \protect \cite{Kennicutt98b} and include a conversion factor of 1.7 to adjust to a Chabrier IMF. The SFR$_{\rm H\alpha}$ values are extinction corrected using an average extinction of dusty, star forming ULIRGs of $\rm A_v=2.9\pm0.5$ \protect \citep{Takata06}. The SFR$_{\rm 1.4GHz}$ values are calculated from the radio fluxes using the far infrared-radio correlation \protect \citep{Ivison10}. The SFR$_{\rm FIR}$ values are calculated from the SED-derived far-infrared luminosities. $\rm V_{\rm obs}$ is the difference between the minimum and maximum velocities measured across the H$\alpha$ velocity field (taken as the 5th and 95th percentiles of the velocity distribution).  We quote the properties of the whole SMMJ0217$-$0503 system as well as the two clear components. The properties of SMMJ0217$-$0503b assume a double Gaussian fit to the disk-like H$\alpha$ emission line profile. The broad line fit to H$\alpha$ of SMMJ0217$-$0503 and SMMJ0217$-$0503a (the AGN component) covers H$\alpha$ and the blended [NII] doublet,  therefore it is not possible to measure the [NII] flux. Caution should be taken with the SFR$_{\rm 1.4GHz}$ value for RGJ0331$-$2739 since the high radio luminosity suggests the radio flux is dominated by the AGN and therefore gives an unphysically high SFR using the far-infrared-radio correlation. The ratios of [NII]/H$\alpha$ we quote include the flux from the 6583$\mu m$ [NII] line only since only this line is included when classing systems as being potentially AGN dominated using the comparison to the data in \protect \cite{Kewley06}. $^1$AGN candidate and $^2$AGN as discussed in Section 3.3. }
\label{tab:basic}
\end{table*}

\subsection{Spatially resolved spectra}
We use the spatially resolved H$\alpha$ emission line maps from our NIFS and SINFONI IFU observations to estimate the distribution of star-formation and dynamics within the galaxies.  After constructing the datacubes, we proceed to fit the H$\alpha$ emission line in each pixel.  The spectra were averaged over 3$\times$3 spatial pixels ($0.15''\times0.15''$ for the NIFS data and $0.375''\times0.375''$ for SINFONI), increasing this up to 5$\times$5 pixels ($0.25''\times0.25''$ for the NIFS data and $0.625''\times0.625''$ for SINFONI) if the signal was too low to give a sufficiently high S/N.  In regions where this averaging process still failed to give an adequate S/N, no fit was made.  Using a continuum fit, we required a minimum S/N of three to detect the lines, and when this criterion is met, we fit the H$\alpha$ emission line with a Gaussian profile allowing the normalization, central wavelength and width to vary. 
  
\begin{figure*}
\centering
\includegraphics[width=0.95\textwidth]{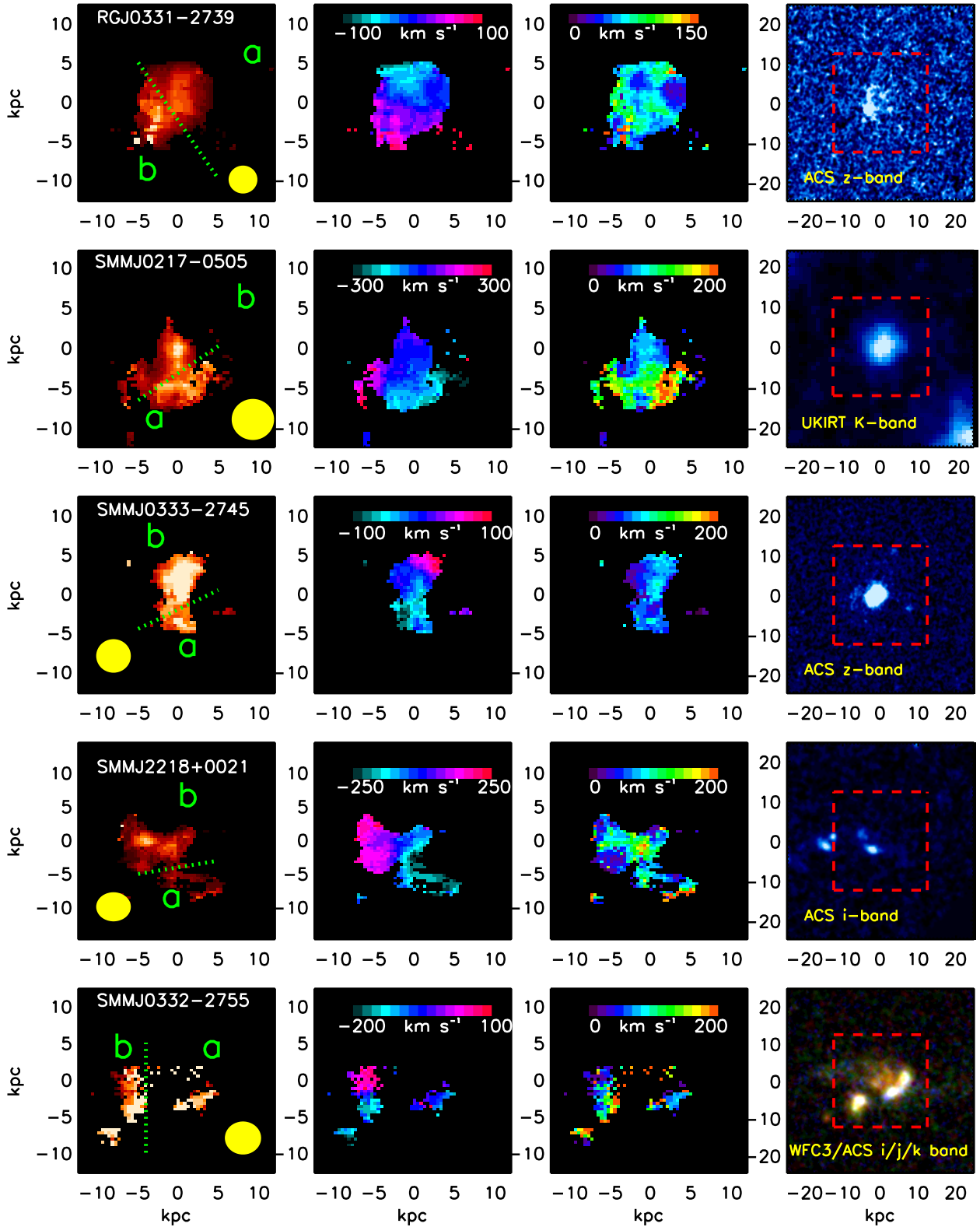}
\caption{Maps of the properties of the H$\alpha$ emission lines for each of the 5 sources observed with NIFS. {\it Left to right}: The H$\alpha$ intensity, H$\alpha$ velocity and H$\alpha$ velocity dispersion maps, and the available imaging (we show HST imaging if it is available otherwise we display the ground based K band imaging). For SMMJ0332$-$2755 imaging in the $i$,$j$ and $k$ bands are available therefore a 3 colour image is displayed. The boxes marked by the red dashed lines represent the field of view of NIFS on the imaging. {\it Top to bottom}: RGJ0331$-$2739, SMMJ0217$-$0505, SMMJ0333$-$2745, SMMJ2218+0021, SMMJ0332$-$2755. The green dotted lines mark the division into components, labelled as `a' and `b'. The yellow circles indicate the FWHM of the seeing dominated PSF from observations of the standard stars. This estimate from the standard stars is, however, likely better than what is achieved during the long science exposures.}
\label{fig:maps}
\end{figure*}

\begin{figure*}
\centering
\includegraphics[width=0.95\textwidth]{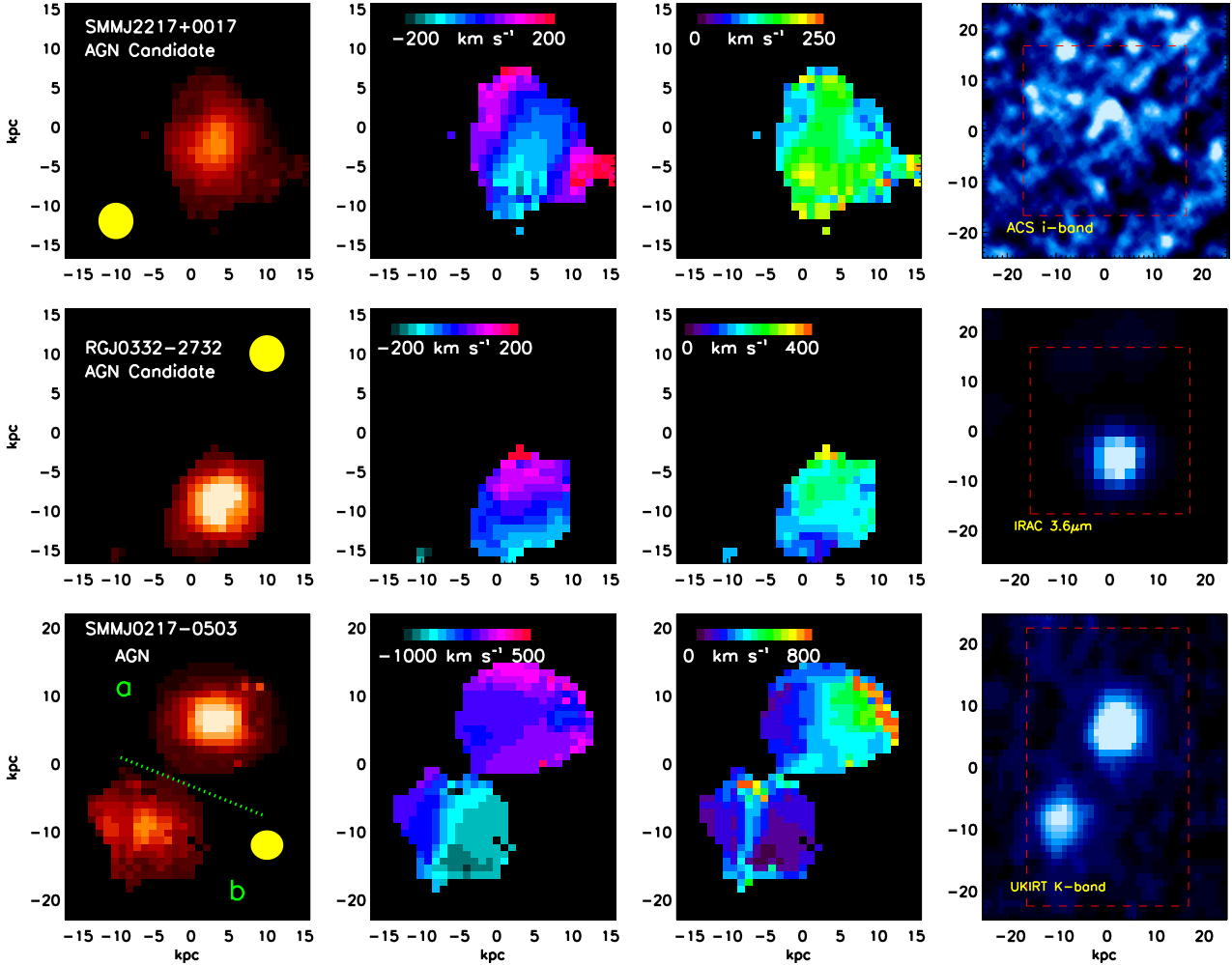}
\caption{Maps of the properties of the H$\alpha$ emission lines for each of the 3 sources observed with SINFONI with strong H$\alpha$ detections. {\it Left to right}: The H$\alpha$ intensity,  H$\alpha$ velocity and H$\alpha$ velocity dispersion maps, and the available imaging (we show HST imaging if it is available otherwise we display the ground based K band imaging or the IRAC 3.6$\mu$m channel map.). The boxes marked by the red dashed lines represent the field of view of SINFONI on the imaging.  {\it Top to bottom}: SMMJ2217+0017, RGJ0332$-$2732 and SMMJ0217$-$0503. The green dotted line mark the division into components for SMMJ0217$-$0503. The yellow circles indicate the FWHM of the seeing dominated PSF from observations of the standard stars. This estimate from the standard stars is, however, likely better than what is achieved during the long science exposures.}
\label{fig:maps_sinf}
\end{figure*}

We simultaneously fit a double Gaussian to the [NII] lines with the fit to H$\alpha$. If the improvement in the $\chi^2$ of this fit, compared to the single Gaussian fit to H$\alpha$, is such that $\Delta\chi^2>9$ (corresponding to 3$\sigma$) then the  H$\alpha$ + [NII] doublet fit is used instead. To calculate the error in the line parameters, we alter the fit parameters one at a time and allowing the other parameters to minimise until $\Delta\chi^2$=1, corresponding to a formal 1$\sigma$ error. Figs. \ref{fig:maps} and \ref{fig:maps_sinf} show the results from the linefitting; 2-dimensional maps showing the H$\alpha$ intensity, velocity and velocity dispersion fields. The yellow circles represent the FWHM of the standard stars showing an estimate of the seeing dominated PSF. This estimate is from shorter exposures ($\sim$10s) than the science exposures and therefore underestimates a possible PSF broadening in the long exposure science observations. We use the H$\alpha$ intensity maps to measure the half-light radii, $\rm r_{1/2}$, by summing the H$\alpha$ flux within increasing circular apertures, centred on the middle pixel of the H$\alpha$ intensity distribution, until the flux enclosed is equal to half the total flux. The $\rm r_{1/2}$ values are then corrected for the PSF.

Figs. \ref{fig:maps} and \ref{fig:maps_sinf} show that in all sources the H$\alpha$ distribution is extended on scales $>$5\,kpc with an average H$\alpha$ half-light radius of $\rm \langle r_{1/2}\rangle=3.7\pm0.8$\,kpc. We compare our H$\alpha$ $\rm r_{1/2}$ measurements to molecular gas sizes in SMGs. \cite{Tacconi06} find a median $\rm r_{1/2}$  of $<$2$\pm$0.8\,kpc\footnote{All given measurements of FWHM and linear diameter are converted to equivalent half-light radius by dividing the diameters by 2.} in their sample of SMGs observed in CO which is in agreement with additional CO measurements from \cite{Engel10} (median equivalent radius of 2.2$\pm$1.4\,kpc) despite the fact that these measurements cover a range of high-J CO transitions possibly tracing different distributions of the gas. These CO sizes are also comparable to the 1.4\,GHz radio sizes for SMGs found by \cite{Chapman04b} and \cite{Biggs08} (median equivalent radius of 1.6$\pm$0.4\,kpc). These compact sizes are lower than the average half-light radius we measure in the H$\alpha$ sample (3.7$\pm$0.8\,kpc) suggesting that the H$\alpha$ observations trace more of the gas distribution in the SMGs. However we do not have CO line data and H$\alpha$ observations in the same sample of SMGs therefore comparing the two samples can only be done statistically.

All sources, except SMMJ2217+0017 and RGJ0332$-$2732, exhibit multiple peaks in the H$\alpha$ intensity distributions with two sources showing clear separations between multiple components (SMMJ0332$-$2755 and SMMJ0217$-$0503). The velocity and velocity dispersion fields are not smooth and do not show evidence for disk-like profiles which is further addressed in Section \ref{sec:kinemetry}.

\subsection{AGN identification}
\label{sec:AGN}

Previous IFU studies of SMGs have mainly targeted galaxies showing strong AGN in their H$\alpha$ spectra \citep{Swinbank05,Swinbank06,Menendez-Delmestre12}, whereas the majority of our sample do not display clear H$\alpha$ AGN signatures. Indeed, the sub-mm and/or far-infrared detections of all the sources in our sample provide a baseline of evidence that they are dominated by star formation (implied $\sim$1000\,M$_\odot$/yr). In SMGs, AGN are found to only contribute at low levels ($<$20\%) to the bolometric output  \citep{Alexander04}. However, even if an AGN is not dominating the bolometric output it may still affect the H$\alpha$ distribution and dynamics, being energetic enough to drive ionized gas over scales of a few kpc (\citealt{Nesvadba08}; Harrison et al.\ in prep.).

Two of our sources (RGJ0332$-$2732 and SMMJ2217+0017) have  high ratios of [NII]/H$\alpha$ in their integrated spectra (Fig. \ref{fig:spectra}), indicating that these sources may host an AGN (a limit of [NII]/H$\alpha >$0.7 is typically adopted for AGN dominated systems \citep{Kewley06}). Given the high SFRs, it is also possible that the high ratios are caused by shock ionization from a galactic wind, as opposed to an AGN ionization field, raising the observed value of [NII]/H$\alpha$ \citep{vanDokkum05}. The detection of H$\alpha$ extended over a large scales ($\sim$10\,kpc), with no obvious outflow kinematics, further suggests these systems may still be star formation dominated in their H$\alpha$ kinematics.

We expect the high [NII]/H$\alpha$ ratio from AGN excitation would typically be concentrated to the inner few kpc. We search the off-nuclear regions of RGJ0332$-$2732 and SMMJ2217+0017 for gradients in the [NII]/H$\alpha$ excitation, finding that SMMJ2217+0017 shows very concentrated [NII] ([NII]/H$\alpha$=0.7$\pm$0.1), with low [NII]/H$\alpha$ ([NII]/H$\alpha$=0.3$\pm$0.1) in regions around the nucleus, likely indicating a compact AGN core surrounded by star formation. 

RGJ0332$-$2732, by contrast, shows continued strong [NII]/H$\alpha$ outside the central nucleus and observations of the [OIII], H$\beta$ emission indicate photoionization by an AGN also (Harrison et al. \ in prep.). The ratio of [NII]/H$\alpha$ is higher in the northern region of this source ([NII]/H$\alpha$=0.9$\pm$0.1 rather than [NII]/H$\alpha$=0.7$\pm$0.1 in the southern region). It is therefore possible that the AGN may be ionising the gas to the north only and the remaining extended H$\alpha$ dynamics we observe are unaffected by AGN activity. The [NII] velocity field follows the same galaxy-wide rotation pattern as the H$\alpha$. However, since there is evidence for the velocity fields in RGJ0332$-$2732 and SMMJ2217+0017 to represent gas which is photoionized by an AGN, given the high [NII]/H$\alpha$ values, we mark these sources as an `AGN candidates' in the analysis.

The high radio flux of RGJ0331$-$2739 (S$\rm_{1.4GHz}$=965$\pm$16\,$\mu$Jy) suggests the presence of an AGN. Although it is luminous in the far-infrared as well, the radio luminosity exceeds that expected from the far-infrared-radio relation \citep{Ivison10} by a factor 29. We do not see evidence for any AGN in the H$\alpha$, [NII] fields, and the AGN-radio core could be very extincted, buried in the system. It is clear from the velocity fields of RGJ0331$-$2739 (Fig. \ref{fig:maps} and Section \ref{sec:kinemetry}) that merger kinematics appear to dominate over any possible AGN-driven properties, and the [NII]/H$\alpha$ ratio lies comfortably within that expected for a $z\sim2$ star forming galaxy \citep{Erb06}. We therefore do not consider this source as an AGN in the remaining analysis for purposes of analyzing its nebular line properties.

SMMJ0217$-$0503 is a particularly complex system. It is clear in Fig. \ref{fig:maps_sinf} that SMMJ0217$-$0503 consists of two spatially separated merging galaxies, with integrated spectra for the individual components shown in Fig. \ref{fig:spectra}. The northern source (SMMJ0217$-$0503a) reveals a broad H$\alpha$ line indicating a dominant AGN. The southern source (SMMJ0217$-$0503b) appears to be star formation dominated (low [NII]/H$\alpha$) with a disk-like velocity field and a double peaked H$\alpha$ line profile in the integrated spectrum (Fig.\ref{fig:spectra}). However, both components in this$\sim$20\,kpc-separated major merger appear to be far-infrared-luminous (e.g., Bothwell et al.\ in prep.; Alaghband-Zadeh et al.\ in prep). Although we measure the properties of these two sources in SMMJ0217$-$0503 separately in the following analyses, we consider the whole system as a major merger. We note however that the large velocity offset of the two components (670$\pm$70\,kms$^{-1}$) implies that it is possible that this is a flyby system rather than a merger, however this is dependent on the orientation, configuration and halo mass.

\section{Results}
\label{sec:results}

\subsection{Dynamical Properties}

\begin{figure*}
\includegraphics[width=0.95\textwidth]{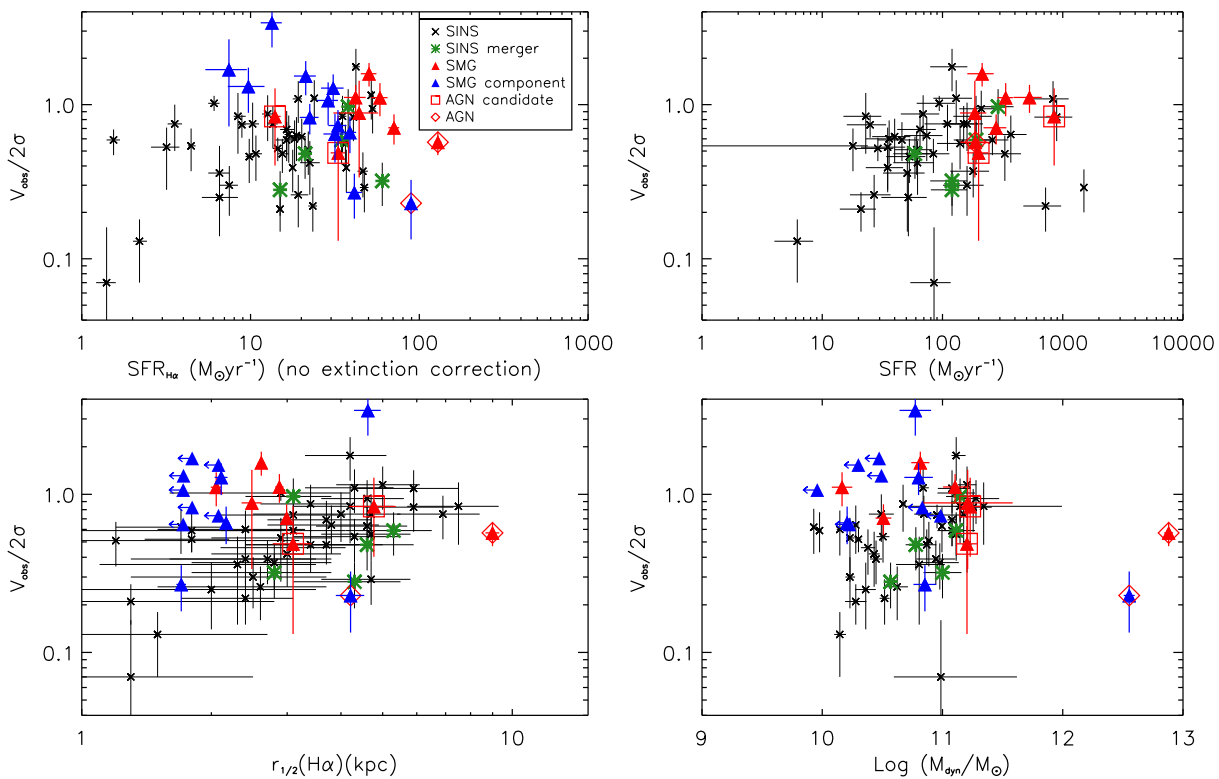}
\caption{Properties of the SMG sample compared to the sample of SINS galaxies. $\rm V_{obs}/2\sigma$ is ratio of half the observed velocity gradient to the integrated velocity dispersion and provides a measure of how rotation or dispersion dominated a system is. We note that in the case of merging systems $\rm V_{obs}/2\sigma$ is not a direct probe of the physical state of the galaxy since it does not represent the velocity gradient across a single galaxy, however it remains useful for the comparison of the kinematic fields to the SINS sample. The error bars represent the 1$\sigma$ uncertainties. {\it Top left:} Relation between the uncorrected H$\alpha$ SFRs and $\rm V_{obs}/2\sigma$. {\it Top right:} Relation between the extinction corrected H$\alpha$ SFRs for the SINS sample, the far-infrared SFRs for the SMGs and $\rm V_{obs}/2\sigma$. {\it Bottom left:} Relation between the H$\alpha$ half light radii and $\rm V_{obs}/2\sigma$. {\it Bottom right:} Relation between the dynamical mass and $\rm V_{obs}/2\sigma$. The masses of the SMG sample represent a useful parameterization in terms of the integrated-velocity dispersion and $\rm r_{1/2}$ for comparison to the SINS sample, but only have a physical representation as a dynamical mass for certain model configurations in virialized systems. We divide SMMJ0217$-$0503 into its two components and mark the northern SMMJ0217$-$0503 source (SMMJ0217$-$0503a) as an `AGN' (open red diamond) and the southern disk-like component as a SMG. We plot the properties of the whole SMMJ0217$-$0503 system also marked as an `AGN'.  We highlight RGJ0332$-$2732 and SMMJ0332$-$2732 as `candidate AGN' (open red squares) since there is some evidence that the velocity fields may be partially affected by AGN given their high values of [NII]/H$\alpha$ (Section 3.3).}
\label{fig:all}
\end{figure*}

In order to test whether the SMGs are distinct in their basic dynamical properties from other galaxy populations, we compare the kinematics of our SMGs to other high-redshift star-forming galaxies studied in previous surveys.  In particular, we calculate the ratio of the maximum velocity across the source to average velocity dispersion, which provides an indicator of the extent to which a system is rotation or dispersion dominated, $\rm V_{obs}/2\sigma$. Here, $\rm V_{obs}$ is the difference between the maximum and minimum velocities derived across the spatial extent of the galaxies, taken as the 5th and 95th percentiles in the velocity distributions determined from the H$\alpha$ line-fitting, while $\sigma$ is the integrated galaxy velocity dispersion measure derived from the integrated spectra. In Fig. \ref{fig:all}, we compare our results to the star-forming galaxies from \cite{Forster-Schreiber09}, UV/optically selected galaxies in a similar redshift range to our sample of SMGs (the SINS sample). The SINS sample contains a wide variety of sources and hereafter comparing to this SINS sample refers to the core majority of the SINS sample of turbulent disk-like systems. We note that the integrated velocity dispersions will be larger than the average dispersions measured in the maps (Figs. \ref{fig:maps} and \ref{fig:maps_sinf}) since the integrated values contain a non-negligible contribution from the range of velocities within the sources. We compare the average velocity dispersion in the maps to the integrated value finding that, on average, the integrated values are $\sim$20\% larger than the average in the dispersion maps. However, since \cite{Forster-Schreiber09} use the integrated velocity dispersion in their calculation of $\rm V_{obs}/2\sigma$ we also use the integrated value to enable a consistent comparison.

Fig. \ref{fig:all} shows $\rm V_{obs}/2\sigma$ plotted against the SFRs, H$\alpha$ half light radii and the dynamical masses. The average $\rm V_{obs}/2\sigma$ of the SMGs is 0.9$\pm$0.1  which is higher than the average of the SINS sample; $\langle \rm V_{obs}/2\sigma\rangle=$0.60$\pm$0.05. However the high values of $\rm V_{obs}/2\sigma$ are due to large velocity gradients within the systems rather than especially low velocity dispersions. The average $\rm V_{obs}$ of our SMGs is 330$\pm$80\,kms$^{-1}$ compared to 160$\pm$20\,kms$^{-1}$ for the SINS sample whereas the average integrated velocity dispersion of the SMGs (excluding the AGN,  SMMJ0217$-$0503) is 130$\pm$\,20kms$^{-1}$ which is consistent with the average for the SINS sample of 130$\pm$7\,kms$^{-1}$. The large velocity gradients in the SMGs may be due to the minimum and maximum velocities coming from different components of a merger system. Indeed, the systems which do not have clear sub-components in Fig.\ref{fig:maps_sinf} (SMMJ2217+0017 and RGJ0332$-$2732) have the lowest values of $\rm V_{obs}/2\sigma$, along with the broad-line AGN component SMMJ0217$-$0503a. The largest value of $\rm V_{obs}$ is in the southern component of SMMJ0217$-$0503b  suggesting a sharp velocity gradient through the disk or possibly a large velocity offset between merging components. When considering merging systems, $\rm V_{obs}/2\sigma$ does not represent the velocity gradient across a single galaxy, and is no longer a direct probe of the physical state of a turbulent disk. However, it remains a useful comparison of the kinematic fields of our SMGs to the SINS galaxies as we attempt to ascertain how the SMGs differ from typical $z\sim2$ disk galaxies.

The average SFR$\rm _{H\alpha}$ of the SMG sample (41$\pm$7\,M$_{\odot}$\,yr$^{-1}$) is somewhat higher than (but within the range of)  the SINS sample  (17$\pm$2\,M$_{\odot}$\,yr$^{-1}$). Correcting the SMGs for the effects of extinction using $\rm A_v =2.9\pm0.5$, the average $\rm A_v$  calculated from a sample of SMGs by \cite{Takata06}, raises the SFR values to approximately align with the $\rm SFR_{FIR}$ values. In the top right-hand plot of Fig. \ref{fig:all} we compare the $\rm SFR_{FIR}$ values of the SMGs to the extinction corrected SFR$\rm _{H\alpha}$ values of the SINS sample since we do not have individual extinction values for the SMGs. We find that the SMGs have higher SFRs, with $\langle \rm SFR_{FIR} \rangle = 370\pm90\,M_{\odot}\,yr^{-1}$, than the SINS sample which has $\langle \rm SFR_{H\alpha} \rangle = 165\pm35\,M_{\odot}\,yr^{-1}$. The value of $\rm A_v$ used in the SINS sample correction comes from SED fitting and also takes into account the fact that Balmer emission is more attenuated than starlight without a direct measurement of this extra attenuation therefore there are significant uncertainties in this correction factor.

The H$\alpha$ half-light radius measurements for the SMGs span a narrower range than the SINS galaxies, with $\rm \langle r_{1/2}\rangle=3.7\pm0.8$\,kpc.  The SINS sample have increasing $\rm V_{obs}/2\sigma$ with $\rm r_{1/2}$ suggesting that as the size of the galaxies increases the galaxies become more rotation dominated. Alternatively, as the sizes increase the velocity differences also increase perhaps due to the velocities having larger ranges when they span larger systems since they lie in more massive halos, however when considering merging systems the size is defined by the merger state rather than the halo mass. This offers an explanation for why we do not observe the strong relationship between $\rm V_{obs}/2\sigma$ and  $\rm r_{1/2}$ in the SMG sample, since we interpret these systems to be mergers (Section \ref{sec:kinemetry}).

The SMG H$\alpha$ morphologies and velocity structures clearly show disturbed and complex dynamics, which motivates using a dynamical mass calculation assuming the systems are not dominated by rotation. We therefore use a virial mass approximation, Equation \ref{eq:mddisp}, which is the method used to calculate dynamical masses for the dispersion-dominated galaxies of the SINS sample, stated by \cite{Forster-Schreiber09} as being suitable for a wide range of galaxy mass distributions \citep{Binney08}.

\begin{equation}
\rm M_{dyn} = \frac{6.7 \sigma^2r_{\rm 1/2}}{G}
\label{eq:mddisp}
\end{equation}

Since we interpret the SMGs as being merger systems (Section \ref{sec:kinemetry}) and the velocity dispersions may be driven by the star formation or AGN (Section \ref{sec:sigsfr}), the virial assumption may not be valid. However, this `mass' provides a useful comparison between the SMGs and SINS galaxies in terms of the integrated velocity dispersion-$\rm r_{1/2}$ parameter space covered. The masses determined from Equation \ref{eq:mddisp} of the SMGs span a range from $\rm 1.4\times10^{10}$\,M$_\odot$ to $\rm 1.6\times10^{11}$\,M$_\odot$. This does not include the three AGN candidates, which have systematically larger line widths, since the  masses calculated using this method may simply reflect the gas motions close to the AGN. Our non-AGN SMG sample shows a similar range to that found in the SINS galaxies, and thus the average velocity dispersions of the SMGs do not obviously distinguish them as more massive, however the masses are further investigated in Section \ref{sec:masses}. The interpolated restframe $H$-band luminosities of our SMG sample suggest they have similar stellar masses to the SMGs studied in \cite{Hainline11} (median $\rm \sim7\times10^{10}\,M_{\odot}$). This places the stellar masses of the SMGs in the upper third of the estimated stellar masses of the SINS sample which supports the possibility that these ULIRGs may be formed from SINS-type galaxies merging.

To quantify the difference of the two samples in these quantities we apply a Kolmogorov-Smirnov test (including only the non-AGN component of SMMJ0217$-$0503). We derive a 40\% probability that the SINS and SMG samples are drawn from the same distribution in terms of $\rm r_{1/2}$ and 96\% for $\rm M_{dyn}$ suggesting that the SMGs are not significantly different to the SINS sample in terms of these properties. The probability that the SMGs' $\rm SFR_{FIR}$ are drawn from the same distribution as the dust-corrected SFR$\rm_{H\alpha}$ in the SINS sample is 0.02\%, while for $\rm V_{obs}/2\sigma$ the probability is 2\%, suggesting a difference in these quantities between our SMG sample and SINS galaxies.

We also consider how the galaxies in the SINS sample classified as mergers compare to our SMGs. The SINS merger subsample are classified as mergers using kinemetry, the method we use in Section \ref{sec:kinemetry}. Particularly in $\rm V_{obs}/2\sigma$ and SFR, the SMGS are distinct from  the SINS merger subsample, which may be indicative of the lower luminosity mergers found in the SINS galaxies.

Despite the large difference in their SFRs, the SINS galaxies and SMGs can not be clearly differentiated simply from basic measures of their velocity difference, velocity dispersion or half-light radius, and so more complex dynamical modelling is required. The SMGs could still represent a mixture of disks and mergers in the same way the SINS sample does, with SMGs being scaled up and more clumpy versions of local disks as seen in the simulations detailed by \cite{Dave10}.  When we consider this further in Section \ref{sec:kinemetry}, where the levels of asymmetry in the velocity and dispersion fields are measured, we find strong evidence that the dynamics of all the SMGs are more consistent with being mergers.

\subsection{Evidence for Mergers and Kinemetry analysis}
\label{sec:kinemetry}

The intensity maps of Fig. \ref{fig:maps} and \ref{fig:maps_sinf} show the distribution of star formation within the sources via the H$\alpha$ morphologies. Six of the sources exhibit clear multiple peaks in H$\alpha$ intensity, three of which match multiple peaks in imaging, which could in some cases be a result of structured dust within the sources obscuring regions of star formation resulting in  clumpy H$\alpha$ morphology \citep{Swinbank10}. The multiple peaks could also be distinct objects with are in the process of merging. The velocity maps show large ranges in the velocities of the systems (typically 400\,kms$^{-1}$ peak to peak) and the dispersion maps show disturbed motion within all the sources, providing initial evidence that most of the SMGs could be merger systems. Indeed only RGJ0332$-$2732 (an AGN candidate) exhibits a smooth velocity and dispersion field. 

To better quantify the dynamics, we attempt to fit the two-dimensional velocity field using simple disk models. We parametrise the two-dimensional velocity field using an arctan function of the form $\rm V(r) = \frac{2}{\pi} V_c arctan(r/r_t)$ where $\rm V_c$ is the asymptotic circular velocity and $\rm r_t$ is the radius at which the rotation curve turns over \citep{Courteau97}.  The model has six free parameters, disk centre [x/y], peak rotational velocity, V$_{\rm c}$, r$_{\rm t}$, disk inclination, (\emph{i}) and sky position angle.  To find the best fit, we use a genetic algorithm with 10,000 possible random fits in each generation, and calculate the $\chi^2$ of each fit. Each parameter is then adjusted towards the best fit parameter and the process is repeated until all the values have converged (and we demand $>$10 generations have passed before testing for convergence). Unsurprisingly, given the complex dynamics seen in Figs. \ref{fig:maps} and \ref{fig:maps_sinf}, simple disk models do not provide an adequate description of the data, with average differences between the best-fit model and the data of  40$\pm$10\,kms$^{-1}$ and a median reduced $\chi^2$ of 60$\pm$30.  As a control sample,  we also run the same fitting procedure on a sample of disks at high-redshift (Swinbank et al.\ in prep.) finding a median rms of 27$\pm$7\,kms$^{-1}$ and reduced $\chi^2$ values between 5 and 8 which suggests the SMGs are not consistent with being disk-like systems. 

In order to provide a more definitive test of whether the kinematics are rotation versus merger dominated, we can use kinemetry, which was originally developed to search for non-circular dynamical features in the stellar dynamics of local ellipticals \citep{Copin01}, but has been extended to the dynamics of high-z galaxies. 

The  kinemetry  described in \cite{Shapiro08} (employing the code http://www.eso.org/$\sim$dkrajnov/idl/)  uses the method detailed in \cite{Krajnovic06}, and has been used to identify the mergers in the SINS sample \citep{Forster-Schreiber09}. This might allow us to better test whether the dynamics of our SMGs are different from SINS galaxies. The asymmetry is measured by fitting ellipses,  varying the position angle and ellipticity, to the velocity and dispersion fields. The ellipses are fit at increasing radii and the coefficients of the Fourier expansion of each best fit ellipse at each radius are recorded. These coefficients are then used to establish the level of asymmetry level. The velocity field asymmetry is measured as the average over all radii of

\begin{equation}
\rm v_{asym} = \langle \frac{k_{avg,v}}{B_{1,v}} \rangle _r
\end{equation}

where $\rm k_{avg,v}=(k_{2,v}+k_{3,v}+k_{4,v}+k_{5,v})/4$ and $\rm k_{n,v}=\sqrt{A_{n,v}^2 + B_{n,v}^2}$ and where $\rm A_{n,v}$ and $\rm B_{n,v}$ are the coefficients of the Fourier expansion of the best fit ellipse. The velocity dispersion field asymmetry is measured as the average over all radii of

\begin{equation}
\rm \sigma_{asym} = \langle \frac{k_{avg,\sigma}}{B_{1,v}} \rangle _r
\end{equation}

where $\rm k_{avg,\sigma}=(k_{1,\sigma}+k_{2,\sigma}+k_{3,\sigma}+k_{4,\sigma}+k_{5,\sigma})/5$ and $\rm k_{n,\sigma}=\sqrt{A_{n,\sigma}^2 + B_{n,\sigma}^2}$ and where $\rm A_{n,\sigma}$ and $\rm B_{n,\sigma}$ are the coefficients of the Fourier expansion of the best fit ellipse.

The two asymmetry measures are combined together as $\rm K_{asym}=\sqrt{v_{asym}^2 + \sigma_{asym}^2}$. Systems with $\rm K_{asym}>0.5$ are classified as mergers. This limit is determined by running the kinemetry analysis on template merger and disk systems, detailed in \cite{Shapiro08}, to find the two populations separate at $\rm K_{asym} \sim 0.5$.

\begin{figure*}
\includegraphics[width=0.9\textwidth]{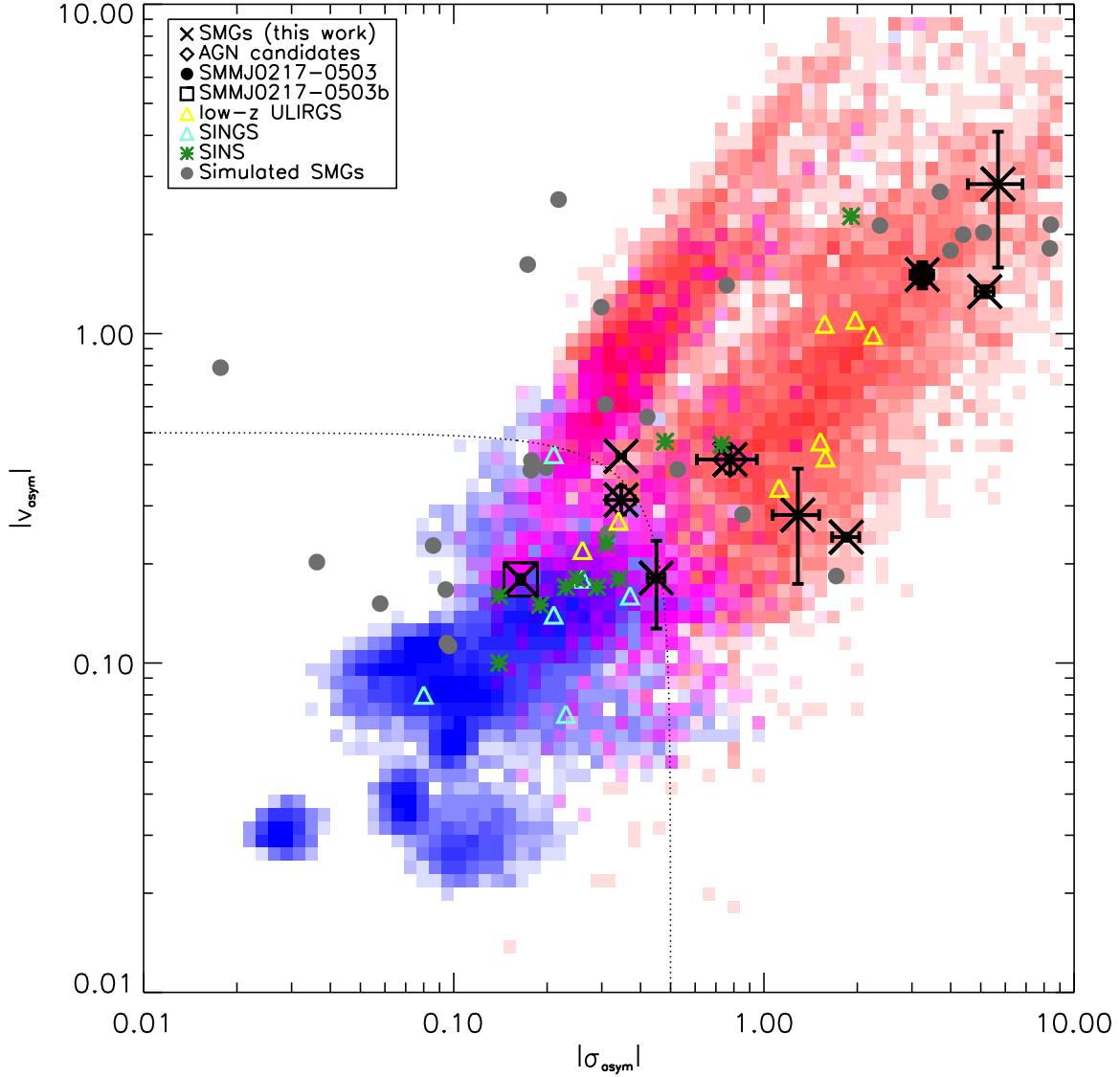}
\caption{The asymmetry measures of the velocity and velocity dispersion fields for the sample of SMGs compared to a number of other samples; the SINGS sample of local spiral galaxies, a sample of low-redshift ULIRGs, the SINS sample of $z\sim$2 star forming galaxies, details of which are given by \protect \cite{Shapiro08}. The background red-blue points are the results of running the analysis on template disks and mergers taken from \protect \cite{Shapiro08}. All of the SMGs lie in the merger region (red background) classifying all of SMGs as mergers. We highlight the AGN candidates, RGJ0332$-$2732 and SMMJ2217+0017 and the whole SMMJ0217$-$0503 system containing the AGN and disk and we also plot the asymmetry values of SMMJ0217$-$0503b only (disk-like component of the merging system). We plot the results of running the kinemetry on a sample of simulated SMGs at $z=2$ finding they represent a range of disks and mergers.}
\label{fig:asym}
\end{figure*}

We fit the centres of the ellipses first and then we re-run the kinemetry 100 times per source varying the centres within the standard deviation of the initial centre fit. The $\rm v_{asym}$ and $\rm \sigma_{asym}$ values we quote in Fig. \ref{fig:asym} are the bootstrapped median and standard deviation of these 100 runs. The $\rm K_{asym}$ values we quote (in Table \ref{tab:basic}) are the bootstrapped median and standard deviation of the 100 runs (rather than using the median $\rm v_{asym}$ and $\rm \sigma_{asym}$ values). Owing to the lower quality of the velocity and dispersion maps of SMMJ0332$-$2755 we bin the maps by 2$\times$2 pixels in order the run the kinemetry. 

In Fig. \ref{fig:asym} we compare the levels of asymmetry in the velocity and dispersion fields to a number of observed and simulated systems, detailed in \cite{Shapiro08}. The SMGs have, on average, higher levels asymmetry that both the low-redshift ULIRG population and the more `normal' star forming galaxies of the SINS sample. The SMGs also do not lie close to the SINGS low-redshift spiral galaxies confirming that the SMGs are very different to other populations of star forming galaxies in terms of their gas dynamics. 

The background red-blue pixels represent the results of running the kinemetry on the template disks and mergers provided by \cite{Shapiro08}. We find all of the systems in our sample of SMGs have $\rm K_{asym}>0.5$ (Table \ref{tab:basic}) and thus we classify all systems as mergers. We also calculate the level of asymmetry in the southern component of SMMJ0217$-$0503 finding that $\rm K_{asym}<0.5$ which suggests that this component is a disk-like source (as implied in Figs. \ref{fig:spectra} and \ref{fig:maps_sinf}), merging with the northern AGN source. Indeed the $\rm K_{asym}$ value for the entire system is 3.9$\pm$0.3 and therefore the whole system is clearly classed as a merger. The sources which are the closest to the disk region of Fig. \ref{fig:asym} are the AGN candidate source (SMMJ2217+0017), RGJ0331$-$2739 and the disk component of SMMJ0217$-$0503 (the southern component in a major merger).  However the $\rm K_{asym}$ values for SMMJ2217+0017 and RGJ0331$-$2739 (using the bootstrapped median of the 100 runs) put these sources in the merger class. Therefore we classify all systems as mergers.

The velocity and dispersion maps cover, on average, a total area of $\sim$160\,kpc$^2$ (at $z=2$). The spatial resolution of the observations is $0.5''$ corresponding to an area of $\sim$16kpc$^2$ therefore on average we have $\sim$10 resolution elements in each map.  We perform an identical analysis on our SMGs to that done on the SINS sample, which have similar spatial extent. We note however that the fit is formally under-constrained and as such, where we find a disk solution, we cannot completely rule out an underlying merger.

We also plot the asymmetry values determined from running the kinemetry on a sample of simulated SMGs at $z=2$ from \cite{Dave10}.  We find these simulated SMGs have a range of asymmetry values forming a sample of both disks and mergers and our observations are not in complete agreement with the simulated systems since we do not observe the range of disks predicted by the simulations. We perform a two dimensional KS test on the model SMGs and observed SMGs to find the probability that both samples are drawn from the same distribution of 0.1 and therefore we cannot rule out that the simulated SMGs represent the observed SMGs. Further, the merger-like kinematics in these simulations does not necessarily imply that the high infrared luminosity is being driven by a merger event, since some of these simulated galaxies are not actively merging.

There is therefore considerable evidence for the SMGs to be merging systems; the multiple peaks in the H$\alpha$ intensity fields, the disturbed velocity and dispersion fields to which 2d disk models could not be fit, the distinction in some integrated quantities between the SMGs and the SINS star forming galaxies, and finally the kinemetry analysis which classes the SMGs as mergers.

\subsection{Sub-components and separations in the  H$\alpha$ intensity field}

Having established that the H$\alpha$ dynamics of the SMGs are likely merger-driven, we next try to identify the components which are in the process of merging, as we have done already for the obviously distinct SMMJ0217$-$0503 a \& b.  We noted before that some of the systems have clearly separated H$\alpha$ intensity peaks, and in more than half the sample these clearly align with structured velocity dispersion peaks, as would be expected for merging, self-gravitating components. The sources are therefore split into components by identifying the minima in the intensity maps, placing the two intensity peaks in different components as shown by the green dotted lines in Figs. \ref{fig:maps} and \ref{fig:maps_sinf}. 

The components of SMMJ2218+0021 and SMMJ0332$-$2755 also match the components seen in the HST imaging shown in Fig. \ref{fig:maps}. We find the division is possible in all sources except the AGN candidates, SMMJ2217+0017  and RGJ0332$-$2732. It should be noted that splitting the sources into components does not necessarily represent two distinct objects which are merging since these systems could be in the later stages of merging where multiple passes have mixed the gas and stars. However, we discuss the component properties as a reference point assuming that we have identified real merging components, acknowledging that these systems may reflect more complex dynamics than a simple merging pair.

A two-dimensional disk model is well fit to the southern component of SMMJ0217$-$0503(b) with a reduced $\chi^2$ of 1.4 and higher order moments (kinemetry) also clearly identify this southern component as a coherent disk (detailed in Section \ref{sec:kinemetry}). However we also note that two regions within SMMJ0217$-$0503b (the eastern and western halves) show different (low) ratios of [NII]/H$\alpha$ (0.44$\pm$0.07 and 0.25$\pm$0.07 for the eastern and western sections respectively). If not dominated by ionization effects, then this suggests that this component may contain regions of varying metallicity and therefore could itself be the product of a recent merger.

We derive $\rm V_{obs}/2\sigma$, $\rm r_{1/2}$, $\rm M_{dyn}$ and SFR$_{\rm H\alpha}$ for the components in the same way as for the whole systems (shown on Fig. \ref{fig:all}).  The components exhibit similar $\rm V_{obs}/2\sigma$ values to the entire merging systems since $\rm V_{obs}$ and $\sigma$ are both smaller in the components ($\langle \rm V_{obs} \rangle$=260$\pm$40\,km\,s$^{-1}$ and $\langle \rm \sigma \rangle$=110$\pm$10\,km\,s$^{-1}$). The comparison of our SMGs' merging components with SINS galaxies suggests it is possible that the merger of two typical SINS isolated disk galaxies may generate a SMG.  However, the properties of the individual components will develop throughout the merger, evolving with each pass and therefore our division into components may not correctly separate the two progenitor merging sources.

\begin{figure}
\includegraphics[width=0.48\textwidth]{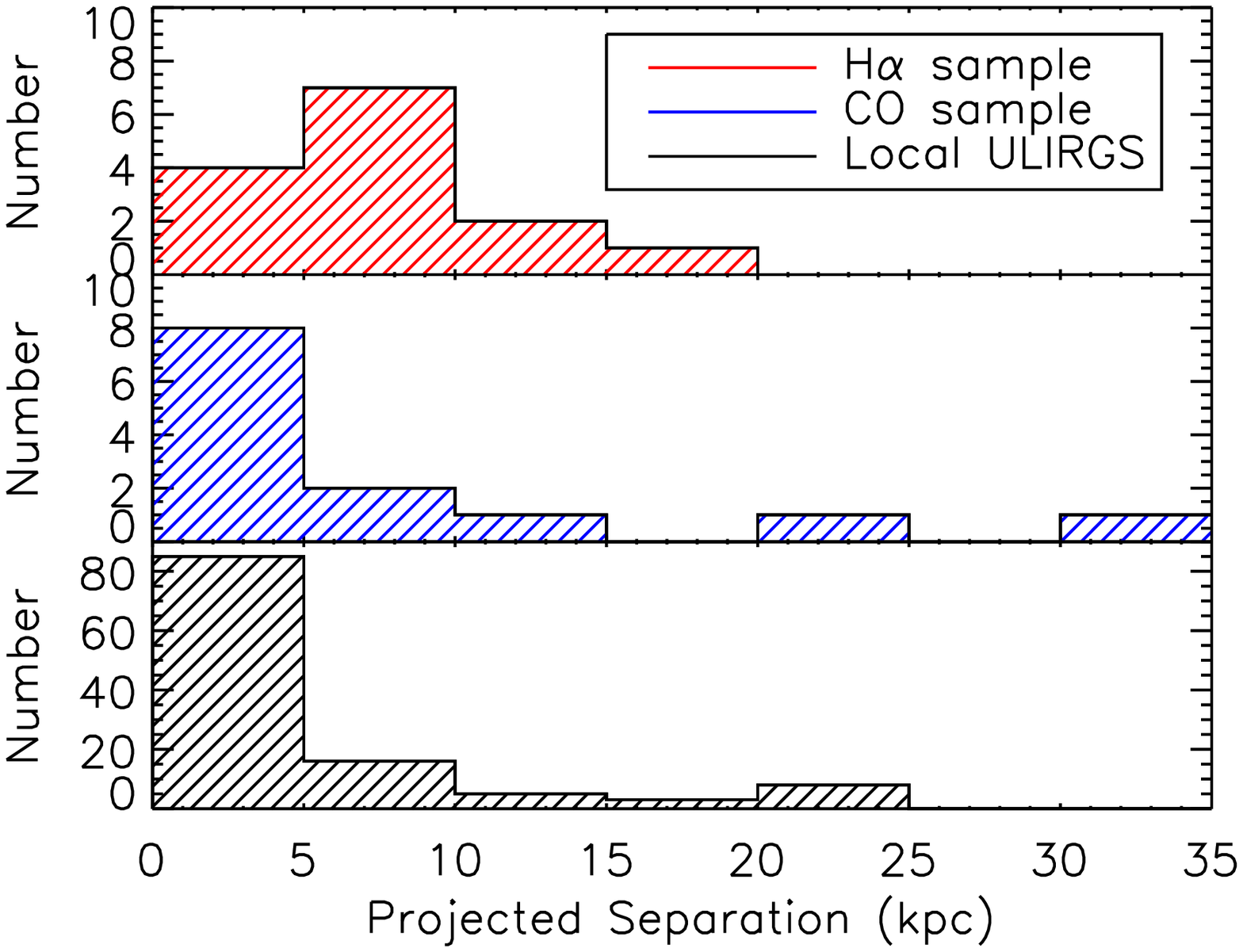}
\caption{Histograms showing the distributions of the projected separations of the SMGs (H$\alpha$ observations in the top plot and CO observations in the middle plot) and a sample of local ULIRGs (bottom). The H$\alpha$ observations combine the NIFS and SINFONI samples with all existing IFU samples detailed in \protect \cite{Swinbank06} and also including \protect \cite{Tecza04}, \protect \cite{Swinbank05} and \protect \cite{Menendez-Delmestre12}. RGJ0332$-$2758 from Casey et al.\ (in prep.) is also included. The H$\alpha$ studied SMGs have larger separations on average than local ULIRGs suggesting different merger stages highlighted by each population. This large separation is not seen in the CO observed SMG sample potentially highlighting the differences in the gas distributions traced by H$\alpha$ and CO.}
\label{fig:sep1}
\end{figure}

In Fig. \ref{fig:sep1}  we show the spatial offsets for our sample, as well as for a sample of SMGs observed in CO, and previous H$\alpha$ studies. These are compared to local  \citep{Veilleux02}  ULIRGs which have multiple components with small separations ($<$4\,kpc apart), a few having components up to 25\,kpc away.  This matches merger simulations \citep{Mihos96} where the large burst of star formation occurs at the final stage of the merger and thus the ULIRGS are expected to be observed in very close configurations. This is however not observed in the SMGs with most of the H$\alpha$ sample having separations greater than 4\,kpc indicating that the ULIRG phase can occur at the early stages of merging in these high-redshift systems. The large separations observed in CO are predicted by hydrodynamical simulations \citep{Hayward11} which explain the 850\,$\mu$m source population.

These 3 samples are complete with the local ULIRGs being cut at $\rm L_{IR}=10^{12}$\,L$_{\odot}$ and both the CO and H$\alpha$ SMG samples spanning a narrow range of $\rm L_{IR}=2-10\times10^{12}$\,L$_{\odot}$. There is however a selection bias since sub-components with r$<$4\,kpc would not be resolved in the H$\alpha$ data. Furthermore, the NIFS H$\alpha$ sample is not sensitive to separations greater than 15\,kpc as the field of view of NIFS is approximately 30\,kpc by 30\,kpc (at $z=2$) whereas the CO SMG sample could detect separations up to $\sim$30\,kpc. The maximum separation detectable by SINFONI is 30\,kpc at $z=2$ due to the larger total field of view of SINFONI, however we nod the source around 4 quadrants of the IFU and therefore the effective field of view is approximately 30\,kpc by 30\,kpc. 

\section{Discussion}
\label{sec:discuss}

The multiple H$\alpha$ peaks, the failed disk-fits to the velocity fields and the kinemetry analysis together provide considerable evidence that these systems are likely to be major mergers. We directly identify multiple components in the majority of the systems and suggest that these are in the process of merging. Our findings agree with those of \cite{Swinbank06} and \cite{Menendez-Delmestre12} who also find multiple-component complex systems in their H$\alpha$ observations of SMGs. These findings align with the conclusions made by \cite{Engel10}, whose CO observations suggest that SMGs are often formed in mergers, and also \cite{Riechers11b} and \cite{Riechers11a} which detail EVLA observations of 3 SMG systems in CO(J=1$-$0) presented as either early stage or late stage mergers. Next we further explore the multi-component nature of the SMG systems and study the observed offsets in the context of the parent dark matter halos.

\subsection{Masses from component properties}
\label{sec:masses}

SMGs have a clustering length $r_0\sim8$\,Mpc \citep{Blain04, Hickox11} consistent with a form of evolution ensuring their properties subsequently match the clustering of $z\sim1$ evolved red galaxies, and finally of galaxy clusters near the present. This should indicate that they occupy the largest dark-matter halos ($\sim 10^{13}$\,M$_\odot$). In order to examine whether the multiple components of the SMGs can be used as a test of the halo mass we investigate the relationship between the projected separations and the velocity offset of the components. 

\begin{figure*}
\includegraphics[width=0.9\textwidth]{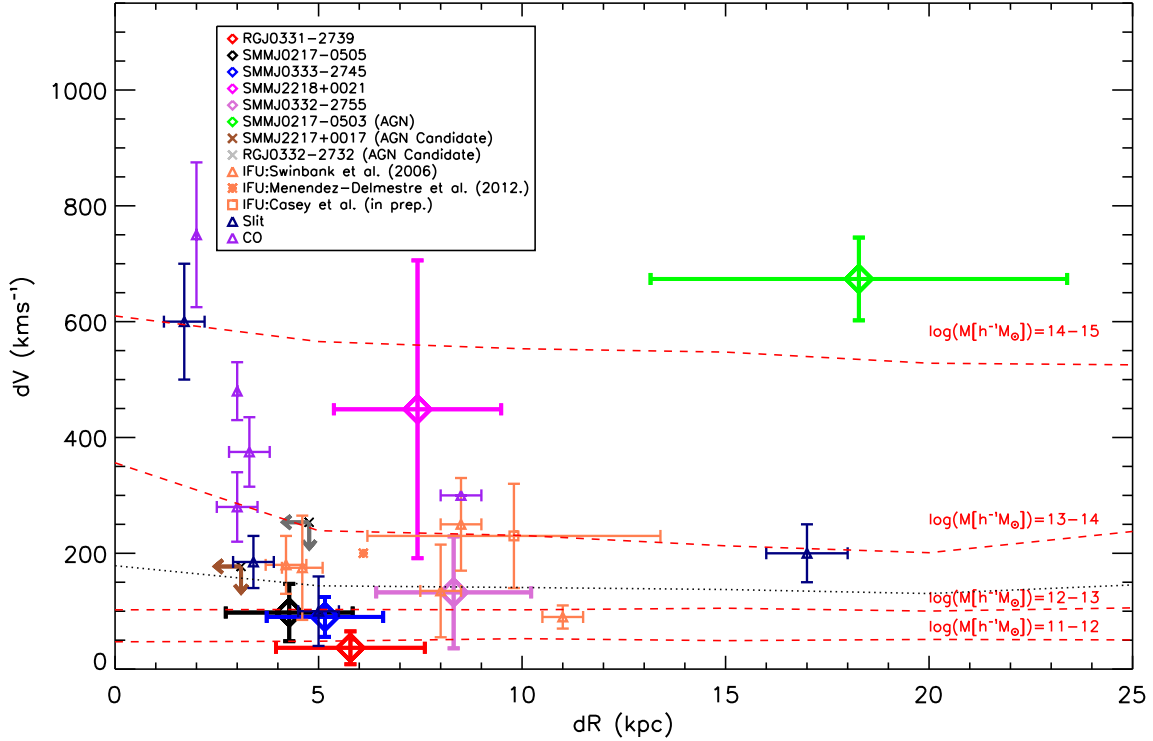}
\caption{Relationship between the offset in position and velocity of the components within the SMGs. Included are the limits on the separations for the two sources which cannot be divided into components (the AGN candidates SMMJ2217+0017 and RGJ0332$-$2732). The outlier is the AGN-disk system of SMMJ0217$-$0503 with a large velocity offset between the two systems. We include the offsets detailed in \protect \cite{Swinbank06} from their IFU observations, and also their compilation of other IFU offsets and those measured from longslit and CO observations \protect \citep{Frayer98,Neri03,Genzel03,Kneib04,Greve05,Tacconi06}. SMMJ123549.44 \protect \citep{Menendez-Delmestre12} and RGJ0332$-$2758  (Casey et al.\ in prep.) are also plotted.  The dashed lines represent the median projected offset between the most-bound sub-halo and the other subhalos within dark matter halos in the quoted mass range from the Millennium Simulation  \protect \citep{Springel05a}. The dotted line represents the interpolation to log($\rm M_{halo}[h^{-1}M_{\odot}])= 12.8^{+0.3}_{-0.5}$, the SMG halo mass measured from clustering \protect \citep{Hickox11}. The observed offsets and the measured offsets from the simulation database are not corrected for viewing orientation of the randomly orientated systems.}
\label{fig:dvdr}
\end{figure*}

Fig. \ref{fig:dvdr} shows the spatial and velocity offsets of the components within each of our systems, combined with those from previous SMG H$\alpha$-IFU and CO observations summarized in \cite{Swinbank06}, along with two additional SMGs with H$\alpha$-IFU measurements in \cite{Menendez-Delmestre12} and Casey et al.\ (in prep.). We also plot the two SMGs in our sample which cannot be divided into components (the AGN candidates SMMJ2217+0017 and RGJ0332$-$2732) with dV, dR limits. The velocity offset for these sources is the difference between the minimum and maximum velocities in the systems ($\rm V_{obs}$) and the spatial offset is $\rm r_{1/2}$, since we assume that if there was a recent merger, it has progressed so that the initial systems are less than $\rm r_{1/2}$ apart and the velocity difference must be below $\rm V_{obs}$. 

Fig. \ref{fig:dvdr} highlights that the CO measurements have generally found significantly smaller component separations than this combined sample of IFU H$\alpha$ maps ($\rm \langle dR\rangle=4\pm1$\,kpc for the CO sample compared to $\rm \langle dR\rangle=10\pm3$\,kpc for the complete H$\alpha$ sample), while the CO velocity offsets extend to high values, with an average of 440$\pm$90\,kms$^{-1}$ compared to 230$\pm$40\,kms$^{-1}$ for the H$\alpha$ sample, consistent with virialised halo models from \cite{Swinbank06}. These clear differences between the CO and H$\alpha$ samples may be indicative of the sensitivity of H$\alpha$ to lower levels of star formation offset from the bright CO region. Indeed, \cite{Swinbank05} and \cite{Menendez-Delmestre12} have clearly demonstrated this to be the case for these two SMGs with both measurements.

Both the spatial and velocity offsets found from the H$\alpha$ components in our SMGs are similar to that of the other SMGs observed in H$\alpha$ with IFUs ($\rm \langle dR\rangle=7\pm1$\,kpc, $\rm \langle dV\rangle=180\pm50$\,kms$^{-1}$ and $\rm \langle dR\rangle=8\pm2$\,kpc, $\rm \langle dV\rangle=200\pm100$\,kms$^{-1}$ for the literature sample and the sample in this work respectively). SMMJ0217$-$0503 is an outlier in the plot with its extremely large spatial and velocity offsets. Given the similarities in overall properties of these various SMGs, we cannot attribute this difference to anything other than small sample variance. However, the wider range of dV found in our expanded SMG sample does make a simple virialised halo model, as adopted in \cite{Swinbank06}, a less satisfactory description of this dataset.

Our goal is to estimate dynamical halo masses for these galaxies, but this is non trivial for what then  may often be unvirialised systems, and so we turn to N-body simulations and extract the average dR and dV for a set of massive halos from the Millennium simulation \citep{Springel05a}. In Fig. \ref{fig:dvdr} we show the projected spatial and velocity offsets for bound sub-halos within dark matter halos of total mass (M). All halos within each mass range are identified and the position and velocity of the most bound sub-halo within each is extracted. All remaining sub-halos are then identified and the spatial (dR) and velocity (dV) offsets between these sub-halos and the most bound sub-halo are calculated for each mass range. We then bin the data in bins of dV and plot the median in each bin to create the tracks for each halo mass. 

The mass range of $\rm 13<log(M[h^{-1}M_{\odot}])<14$ appears to best represent the combined dataset. For comparison, we show the interpolated relationship to  log($\rm M_{halo}[h^{-1}M_{\odot}])= 12.8^{+0.3}_{-0.5}$, the SMG halo mass measured from clustering \citep{Hickox11}, which is lower than the average halo mass we derive. The clustering measurements however come from relatively small samples of SMGs and therefore have large statistical errors. Furthermore, the bias of SMGs to halos is potentially quite complicated \citep{Chapman09}, given the short duration of the burst and the unknown range of environments that they occur in.

The large spatial and velocity offsets of SMMJ0217$-$0503 suggest this system could have a $\rm log(M[h^{-1}M_{\odot}]) \sim 14$ halo which we view along the line of sight velocity, or an even larger halo mass if it is not in this preferred viewing orientation, providing evidence that SMGs may sometimes be beacons for massive proto-clusters of galaxies ($\rm log(M[h^{-1}M_{\odot}])>14$) \citep{Chapman09, Tamura09, Matsuda11}. In the Millennium Simulation database there are 28 halos with masses  $\rm 14<log(M[h^{-1}M_{\odot}])<15$ confirming that these large masses are possible and therefore that the two components of SMMJ0217$-$0503 are likely part of a common halo rather than two separate sources whose projection places them within our field of view. However we note that it is also possible that the SMMJ0217$-$0503 system may represent a flyby rather than a merger owing to the large velocity offset of 670$\pm$70\,kms$^{-1}$.

\subsection{High velocity dispersions and the relationship to SFR density}
\label{sec:sigsfr}

Figs. \ref{fig:maps} and \ref{fig:maps_sinf} clearly show resolved substructure in the H$\alpha$ velocity dispersion maps which often correlate with the intensity fields. We further investigate this relationship and explore the origin of the high velocity dispersions found in our SMGs. Despite the larger $\rm V_{obs}/2\sigma$ in the SMGs compared to the SINS galaxies, the values are significantly lower than observed in local disks ($\rm v/\sigma \sim 2-5$: \citealt{Hunter05}) which have much lower dispersions but comparably large circular velocities. Given our  findings (in Section \ref{sec:kinemetry}) that the asymmetrical velocity and dispersion fields enable us to classify the systems as mergers, it is natural that the high velocity dispersions within our sample may be due to the merging systems, causing the gas to be disturbed.

The SINS galaxies with high velocity dispersions are explored in \cite{Genzel08} within the context of disk-like galaxies displaying turbulent motion. They explain the turbulent, rotating disk structures either using feedback from the star formation driven by supernovae winds \citep{Efstathiou00}, or with accreting gas driving the turbulence as it flows into the disk from the dark halo \citep{Genzel06}. By contrast, the simulations of \cite{Dave10} suggest turbulent motion in massive disk-like galaxies is driven by harassment by smaller satellites and the infall of gas. In the SMGs, the high H$\alpha$ velocity dispersions may also be connected to the star formation itself, rather than the forces acting on the gas from the merger.

We first test whether the relationship between the velocity dispersions and the star formation rates can be explained as being purely a result of the Kennicutt-Schmidt (KS) star formation density - gas relation \citep{Kennicutt98a}:

\begin{equation}
\rm \Sigma_{SFR} = A \Sigma_{gas}^n
\end{equation}

We relate $\rm \Sigma_{gas}$ to the velocity dispersion ($\sigma$) by setting Toomre's Q Criterion to equal to 1, since we assume the gas finds a stable disk configuration in a short time as suggested by hydrodynamical simulations \citep{Springel05b}, where

\begin{equation}
\rm Q = \frac{\kappa \sigma_{gas}}{\pi G \Sigma_{gas}}
\end{equation}

and $\rm \kappa=\sqrt{2}V/r$ \citep{Schaye05} where we take $\rm V=V_{obs}$ and $\rm r=r_{1/2}$. We take $A$ to be $10^{-3.63}$ and $n=1.27$ as derived in \cite{Genzel10}.

\begin{figure*}
\includegraphics[width=0.95\textwidth]{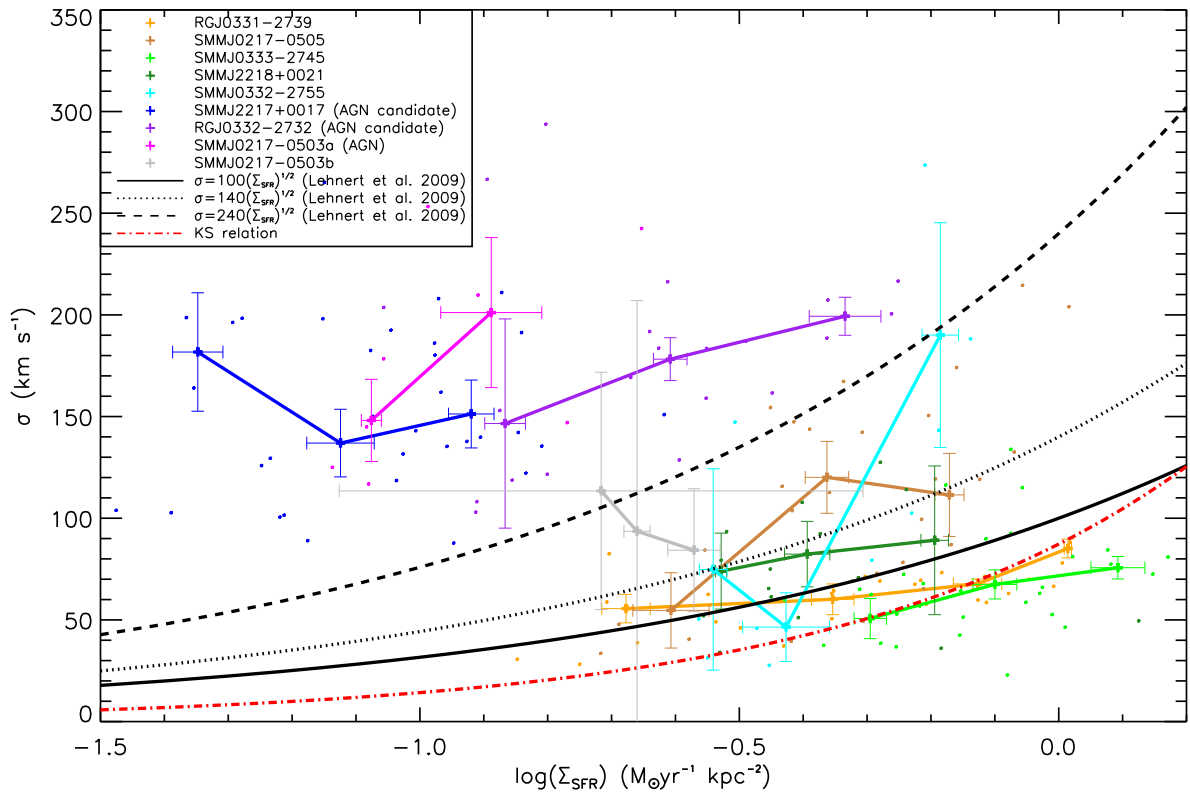}
\caption{Velocity dispersion against the star formation intensity per pixel for each source. The dot-dashed line represents the relation for the case if the velocity dispersion values are purely a result of the Kennicutt-Schmidt (KS) relation. The dashed, dotted and solid lines represent models, detailed in \protect \cite{Lehnert09}, describing the relation between the velocity dispersion and the star formation intensity per pixel if the energy from the star formation is powering the motion of the gas. The three lines assume three efficiencies of the coupling between the input energy by stars and the ISM. The sources follow tracks along the plot despite being offset from each other. The SFRs are not extinction corrected. All sources are smoothed the match the seeing of 0.6'' and then binned so that each pixel is 0.25'' by 0.25'' except for SMMJ0217$-$0503b. Since we observe the disk-like double-peaked H$\alpha$ profile in SMMJ0217$-$0503b we plot the velocity dispersion and star formation intensity in three distinct regions across the source instead of using the results from the line-fitting procedure which blend the double peak as a single line}. 
\label{fig:sfrd}
\end{figure*}

We plot this relation in Fig. \ref{fig:sfrd} to enable a comparison to the observed relations between the H$\alpha$ velocity dispersion per pixel and the star formation intensity. We note first that there are several selection effects at work. There are sections of the $\rm \sigma-\Sigma_{SFR}$ plane which cannot be populated from the observations. The upper left region of Fig. \ref{fig:sfrd} is empty as broad lines with low star formation intensities have too low S/N for Gaussian fitting. There is also a lower limit to the velocity dispersion restricted by the spectral resolution and the lower region ($\sim <50$\,kms$^{-1}$) cannot be populated. 

The high velocity dispersion per pixel values of our three SMGs showing some AGN characteristics could be arising from ionization of gas in the broad- and narrow-line regions of the AGN itself  and it is possible that the lower H$\alpha$ densities in these sources are due to feedback from the AGN inhibiting the star formation in the regions surrounding the AGN \citep{Nesvadba08}. One SF-dominated SMG, SMMJ0217$-$0503b, which has intrinsically narrow lines, is not sufficiently resolved spatially to separate the double-peaked rotation profile in the central region, and our line-fitting procedure blends this double peak as a single line. We therefore choose three distinct regions within the source and use the integrated spectra of these regions to obtain the velocity dispersion and star formation rates, allowing a fit to the double peak.

Fig. \ref{fig:sfrd} shows that the KS relation does not describe the data well since the majority of the data have higher values of velocity dispersions at a given star formation intensity than predicted by this model. Clearly, both the dust extinction and inclination will have a dramatic effect on the location of these tracks.  An average A$_v$=2.4$\pm$0.6 is required for the SMGs (excluding the AGN candidates) to match this $\rm \Sigma_{SFR}$--$\sigma$ relation, which is compatible with the average implied dust correction derived for SMGs \citep[eg][]{Smail04, Swinbank04, Takata06}, and suggestive that SMGs simply follow the standard KS-law (as suggested may be the case by \citealt{Ivison11}).

Since we cannot constrain the dust extinction or inclination angle from our data we also consider if the velocity dispersions are driven by the star formation itself. We follow the same analysis as carried out in \cite{Lehnert09} when exploring the source of the turbulence in SINS galaxies. The SMGs are comparable to the SINS galaxies, with the majority of the data lying within the range spanned by the relations from \cite{Lehnert09} which represent the plausible ranges in efficiency of the energy transfer from the star formation to the gas. It is therefore possible that it is the star formation which drives the velocity dispersion. However it is also possible that the turbulence is driven by the merging processes which stir up the gas. We require extinction maps to establish the true star formation intensities in order to further test these hypotheses.

\section{Conclusions} 
\label{sec:conc} 
We analyse subarcsecond resolution H$\alpha$ line data, taken using the integral field spectrometers Gemini-NIFS and VLT-SINFONI, of nine SMGs with $2.0<$z$<2.7$. The gas dynamics and morphologies were mapped by tracing the H$\alpha$ line within the eight of the sources. We compare the velocity fields in this sample to a sample of UV/optically selected star forming galaxies at the same redshift (studied in the SINS survey) finding that the SMGs are kinematically distinct, with higher SFRs and larger velocity gradients across the systems. The majority of the SMG sample are not consistent with being disk-like systems with multiple peaks in H$\alpha$ intensity and irregular and turbulent velocity and velocity dispersion fields. This is confirmed from the kinemetry analysis which classes all systems as mergers rather than disks. We identify multiple components in six of the sources further confirming the suggestion that most SMG star bursts are triggered in major mergers \citep{Engel10}.  

We bring together all existing IFU H$\alpha$ observations of SMGs and use the spatial and velocity offsets between the components of the merging systems to infer an average halo mass,  $\rm 13<log(M[h^{-1}M_{\odot}])<14$, for the systems which is higher than the mass predicted from clustering measurements. 

We observe higher values of $\rm V_{obs}/2\sigma$ in the SMG sample compared to the SINS galaxies due to the large $\rm V_{obs}$ values measured across merging components. The velocity dispersion values in the SMGs are larger than observed in local disks.  We explore the origin of this turbulence finding that the disturbed gas motions are consistent with the Kennicutt-Schmidt law and variable extinctions, but might also be driven by the merging torques or even the star formation itself.

Analysing the outputs from the various simulation of SMGs in an identical way to the observations will enable us to better determine which situation best describes the trigger for the ULIRG phases in these galaxies. Initial evidence from the kinemetry analysis of the simulated SMGs suggests that the simulated systems are a mixture of disks and mergers. By better understanding the triggers of the ULIRG phases we can also probe the build of mass in some of the most massive galaxies in the Universe and investigate if these SMGs are the progenitors of the massive elliptical galaxies observed today \citep{Hainline11, Gonzalez11}.

\section*{Acknowledgements}
This paper is based on observations made with Gemini-North under program number GN/2010B/42 and ESO Telescopes under program number 087.A-0660(A). We thank Kristen Shapiro for supplying the kinemetry values for template disks and mergers. SA-Z thanks Paul Hewett for his valuable comments and support. SA-Z, CMH and DMA acknowledge the support from STFC. DN acknowledges support from the US National Science Foundation via grant: AST-1009452. AMS acknowledges an STFC Advanced Fellowship.  IRS acknowledges support from STFC and through a Leverhulme Senior Fellowship. This research has made use of data from HerMES project (http://hermes.sussex.ac.uk/). HerMES is a Herschel Key Programme utilising Guaranteed Time from the SPIRE instrument team, ESAC scientists and a mission scientist. HerMES will be described in \cite{Oliver12}. The HerMES data was accessed through the HeDaM database (http://hedam.oamp.fr) operated by CeSAM and hosted by the Laboratoire d'Astrophysique de Marseille.
\bibliographystyle{mn2e}

\bibliography{refs_short}

\end{document}